\documentclass[a4paper,fleqn]{cas-dc}

\usepackage{caption}
\usepackage{float}
\usepackage{subcaption}
\usepackage{siunitx}
\usepackage{pbox}
\usepackage{ragged2e}
\usepackage{multirow}
\usepackage{tabularray}
\usepackage{tabularx}
\usepackage{changepage}
\usepackage{adjustbox}
\usepackage[switch, columnwise]{lineno}

\usepackage[numbers]{natbib}

\def\tsc#1{\csdef{#1}{\textsc{\lowercase{#1}}\xspace}}
\tsc{WGM}
\tsc{QE}
\tsc{EP}
\tsc{PMS}
\tsc{BEC}
\tsc{DE}

\newcommand{\rebuttal}[1]{\textcolor{black}{#1}}

\newcommand\highlightReference[1]{%
  \expandafter\newcommand\csname highlightReference-#1\endcsname{}%
}
\usepackage{xpatch}
\makeatletter
\xapptocmd\@lbibitem{\hlbibitem{#2}}{}{}
\makeatother
\def\hlbibitem#1 #2\par{%
  \expandafter\ifx\csname highlightReference-#1\endcsname\relax
    #2\par
  \else
    \highlight{#2}\par
  \fi
}
\usepackage{color}
\newcommand\highlight[1]{\textcolor{red}{#1}}

\begin{document}

\let\WriteBookmarks\relax
\def\floatpagepagefraction{1}
\def\textpagefraction{.001}
\shorttitle{Influence of the flow split ratio on the position of the main atrial vortex: implications for stasis on the left atrial appendage}
\shortauthors{S. Rodríguez-Aparicio et~al.}

\title [mode = title]{Influence of the flow split ratio on the position of the main atrial vortex: implications for stasis on the left atrial appendage}                     



\author[1]{Sergio Rodríguez-Aparicio}


\address[1]{Departamento de Ingeniería Mecánica, Energética y de los Materiales, Universidad de Extremadura, Avda. Elvas s/n, Badajoz 06006, Spain}

\author[1,2]{ Conrado Ferrera}
\address[2]{Instituto de Computación Científica Avanzada (ICCAEX), Avda. Elvas s/n, Badajoz 06006, Spain}

\author[3]{ María Victoria Millán-Núñez}
\address[3]{Servicio de Cardiología, Hospital Universitario de Badajoz, Avda. Elvas s/n, Badajoz 06006, Spain}

\author[4]{ Javier García García}
\address[4]{Departamento de Ingeniería Energética, Universidad Politécnica de Madrid, Avda. de Ramiro de Maeztu 7, Madrid 28040, Spain}

\author[4]{ Jorge Dueñas-Pamplona}
\cormark[1]
\ead{jorge.duenas.pamplona@upm.es}


\cortext[cor1]{Corresponding author}


\begin{abstract}
\noindent\textbf{Background:} Despite \rebuttal{the} recent advances in computational fluid dynamics (CFD) techniques applied to blood flow within the left atrium (LA), the relationship between atrial geometry, flow patterns\rebuttal{,} and blood stasis within the left atrial appendage (LAA) \rebuttal{remains unclear}.  A better understanding of this relationship would have important clinical implications, as thrombi originating in the LAA are a common cause of stroke in patients with atrial fibrillation (AF). 

\noindent\textbf{Aim:} To identify the most representative atrial flow patterns on a patient-specific basis and study their influence on LAA blood stasis \rebuttal{by} varying the flow split ratio and some common atrial modeling assumptions.

\noindent\textbf{Methods:} \rebuttal{Three} recent techniques were applied to \rebuttal{nine} patient-specific \rebuttal{computational fluid dynamics (CFD)} models of patients with AF: a kinematic atrial model to isolate the influence of wall motion \rebuttal{because of} AF, projection on a universal LAA coordinate system, and quantification of stagnant blood volume (SBV). 

\noindent\textbf{Results:} We identified three different atrial flow patterns \rebuttal{based} on the position of the center of the main circulatory flow. The results also illustrate how atrial flow patterns are highly affected by the flow split ratio, increasing the SBV within the LAA. \rebuttal{As} the flow split ratio is determined by the patient's lying position, the results suggest that the most frequent position adopted while sleeping may have implications \rebuttal{for} the medium\rebuttal{-} and long-term \rebuttal{risks} of stroke.
\end{abstract}




\maketitle

\section{Introduction}\label{sec:intro}

Cardiovascular diseases account for approximately 30\% of all annual deaths worldwide \cite{OMS}. Among \rebuttal{these}, \rebuttal{ischemic} heart disease and stroke were the two leading causes of death \rebuttal{between} the period \rebuttal{1990--2019} \cite{Naghavi24}. Atrial fibrillation (AF) is the most common type of cardiac arrhythmia \cite{LSC21}. Its prevalence is estimated to \rebuttal{reach approximately} 16 million in the United States \rebuttal{by} 2050 \cite{Sanatkhani2023} and 17.9 million in the European Union \rebuttal{by} 2060 \cite{Tsao2023heart}. During an episode of AF, thrombi can originate in the left atrium (LA). These thrombi will cause more than \rebuttal{20\%} of the 18 million yearly ischemic strokes. Furthermore, another \rebuttal{30\%} of ischemic strokes are believed to originate in the LA in individuals with subclinical AF or sinus rhythm \cite{kamel2016atrial}.

AF is typically initiated by an alteration in \rebuttal{the} electrical impulses of pulmonary veins (PV), producing an irregular pattern of atrial contraction \cite{SBKJH20}. These episodes are called paroxysmal when they last briefly or \rebuttal{permanent} when the heart cannot return to sinus rhythm. During these episodes, abnormal \rebuttal{contractions lead} to blood stasis in the left atrial appendage (LAA), increasing the risk of thrombosis. Patients with AF have a five-fold higher risk of stroke than those without AF \cite{Wolf1991}. If AF persists for months or years, it can initiate several mechanisms that can cause adverse atrial remodeling\rebuttal{, including} a variety of tissue changes that alter the biomechanical and electrical function of the myocardium. This remodeling process affects the morphology of the LA/LAA bundle, the recurrence of AF episodes, and increases the risk of stasis in the LAA \cite{boyle2021fibrosis}. This stasis increase is reported even under sinus rhythm \cite{back2023}, as atrial remodeling seems to be correlated with a lower atrial ejection fraction and a higher LA retention ratio. 

The LAA is a cavity within the LA that is a remnant of the embryonic \rebuttal{developmental} stage \cite{Al-Saady1999}. \rebuttal{Their} morphology depends on the individual \cite{lupercio2016left}. Although the exact physiological purpose of \rebuttal{the} LAA is not fully understood, it is believed to act as a contractile reservoir or decompression chamber, depending on the cardiac cycle phase \cite{hoit2014left}. Over time, atrial remodeling can cause LAA enlargement and loss of contractility, \rebuttal{thereby} generating prothrombotic substrate \cite{Al-Saady1999}.

In recent times, the development of increasingly complex computational fluid dynamics (CFD) models has allowed exploration of flow patterns within the heart \cite{Moradi2023}. These models range from those of the entire left heart \cite{mihalef2011patient} to two-cavity models that include the LA and left ventricle (LV) \cite{back2024,bucelli2022mathematical,chnafa2014image, Vedula2015} or to single-cavity models that usually focus on LA or LV \cite{chnafa2014image, Lantz2018a, Otani2016}. 

Most single-cavity models have focused on \rebuttal{the} LV \cite{seo2014effect, vedula2016effect}, but the number of studies \rebuttal{analyzing} LA hemodynamics, particularly stasis within the LAA\rebuttal{, is increasing} \cite{Bosi2018,corti2022impact, duenas2021comprehensive,duran2023pulmonary, Garcia-Isla2018, garcia2021demonstration, Koizumi2015, Masci2019,Otani2016}. Their primary motivation \rebuttal{was} the clinical evidence of the connection between LAA morphology and the risk of thrombosis in patients with AF \cite{DiBiase2012, yamamoto2014complex}. Specifically, these studies focused on the \rebuttal{relationships} between LAA morphology and stasis \cite{duenas2022morphing,Garcia-Isla2018, garcia2021demonstration, Masci2019}, the development of new metrics to quantify this stasis \cite{corti2022impact,Achille2014, duenas2021comprehensive,mussotto2024}, and \rebuttal{a} comparison of common modeling assumptions in LA fluid-dynamic simulations \cite{duenas2021comprehensive, duenas2021boundary,gonzalo2022non,duran2023pulmonary,Mill2021}. Recently, multiphysics models integrating the fluid-structure interaction with the electrophysiological or biomechanical mechanisms related to AF have been proposed \cite{bucelli2022mathematical, corti2022impact, feng2019analysis,gonzalo2022non,mussotto2024}.

The research conducted \rebuttal{thus} far has provided insight into the mechanisms of AF and information on the proper generation of models, boundary conditions, and numerical methods to \rebuttal{accurately} simulate the flow dynamics of the LA. However, the relationship between \rebuttal{the} blood flow and thrombosis within the LAA is complex and multifactorial \cite{boyle2021fibrosis}. This is \rebuttal{because of} the complexity and variability of the LA/LAA bundle, differences in wall movement depending on \rebuttal{the} cardiac condition, paroxysmal nature of atrial fibrillation, and atrial remodeling that occurs in patients with AF over time. 

\rebuttal{One aspect} not considered in most atrial models is the non-Newtonian nature of blood. Blood is generally considered a Newtonian fluid, taking advantage of \rebuttal{its} elevated shear rates and aggregation times \cite{A18, CUDGNG67}. However, \rebuttal{the} LAA blood \rebuttal{velocity is} usually very low in patients with AF. Red blood cells (RBCs) tend to concentrate and aggregate, causing viscosity variations throughout the LAA volume. Recent studies \cite{Gonzalo22,Sanatkhani2023,Zhang2023} have reported that the non-Newtonian effects within LAA should be taken into account, as they increase residence time (RT), viscosity, and thus thrombosis risk. 

Another aspect to \rebuttal{be analyzed} is the PV flow split ratio. The flow split ratio \rebuttal{depended} on the individual. The sum of flow rates entering the left (right) PV \rebuttal{ranged} from 43\% (57\%) to 52\% (48\%), where 47\% (53\%) is the mean value \cite{CTLGF06,Nakaza2022}. This mean value can vary slightly within an individual if measured while \rebuttal{performing} physical exercise \cite{CHTF05}. This variation \rebuttal{was} much more noticeable when the lying position \rebuttal{was} considered \cite{WRAPU19}: left size (59\% LPVs / 41\% RPVs) or right size (37\% LPVs / 63\% RPVs). Furthermore, the flow split ratio variation is considerable after left upper lobectomy in patients with lung cancer \cite{Otani2022,Yi2023}.

\rebuttal{Most} CFD studies \rebuttal{have considered} an equal flow split ratio, resulting in an even flow distribution between the four PVs. Recently, two \rebuttal{studies} have considered the flow split inequality. Lantz \textit{et al.} \cite{Lantz2018a} used four-dimensional cardiovascular magnetic resonance flow imaging (4D Flow CMR) to measure the flow split in two women and one man. The data obtained were compared with 20 flow split variations per patient \rebuttal{to detect} differences in LA flow when variations in the flow split ratio were \rebuttal{considered}. To our knowledge, Durán \textit{et al.} \cite{duran2023pulmonary} is the only work that studies the influence of the flow split ratio on LAA flow. They analyzed the influence of four flow split ratios (40/60, 45/55, and \rebuttal{50/50\%LPV/\%RPV}), \rebuttal{showed} that variations in the flow split \rebuttal{affect} LAA flow patterns and RT. However, the \rebuttal{considered} flow split ratios did not reach the values found when a person lies in the right or left position \cite{WRAPU19}, and more studies are required to fully understand this effect and its implications on \rebuttal{the} stasis within the LAA. This aspect is important \rebuttal{because} adopting a regular sleeping position involves maintaining a certain flow split ratio for many hours, thus \rebuttal{affecting} the patient's stroke risk.

Another challenging \rebuttal{limitation} of CFD atrial studies is the wide diversity of \rebuttal{the} LA and LAA morphologies, making it difficult to compare simulation results from different patients or even between different cardiac conditions for the same patient \cite{bifulco2021computational}. Fortunately, the recently introduced \rebuttal{universal} \rebuttal{l}eft \rebuttal{a}trial \rebuttal{a}ppendage \rebuttal{c}oordinates (ULAAC) allow to project the simulation results \rebuttal{onto} a squared two-dimensional (2D) domain, facilitating visualization for further analysis \cite{duenas2024reduced}. In addition, this technique facilitates dimensionality reduction \cite{balzotti2023reduced, duenas2024reduced}.

As mentioned \rebuttal{previously, it is necessary} to increase the information provided by models \rebuttal{that consider} blood rheology. Moreover, there are gaps in the \rebuttal{literature} when different flow split ratios are considered, and to the best of our knowledge, there is no information on models \rebuttal{that combine} the non-Newtonian nature \rebuttal{of blood} and flow split variation in non-rigid LA geometries. Therefore, \rebuttal{the main objective of this study} was to \rebuttal{better} understand the relationship between LA geometry, flow patterns, and stasis in \rebuttal{the} LAA when the flow split ratio, blood rheology, and wall movement \rebuttal{were} modified. \rebuttal{Nine} patients with AF were imaged and segmented to overcome the limited number of patients in most atrial patient-specific studies. Doppler measurements were available for all patients, allowing both the establishment of boundary conditions and the validation of the simulations. Three recent methodological advances were used in combination with CFD techniques: an LA wall motion model to isolate the effect of atrial wall motion \rebuttal{because of} AF, projection of the results in a ULAAC system to facilitate data visualization and calculation of the stagnant blood volume (SBV). 

Flow analysis allowed the identification of three atrial flow patterns depending on the position of the main atrial vortex. The results suggest that both the flow split ratio and endocardial geometry play an important role in the position of the main atrial vortex, which \rebuttal{simultaneously} has a relevant influence on the formation of \rebuttal{an} SBV within the LAA\rebuttal{,} and thus on the risk of thrombosis. 
As the flow split ratio takes extreme values while lying on the \rebuttal{left and right sides} and the influence of the flow split ratio on the position of the main atrial vortex presents important differences depending on the patient-specific geometry, CFD techniques are once again a tool of great interest to improve the diagnosis and treatment of this patient. By performing a patient-specific simulation, it is easy to determine on which side the patient should preferably sleep to enhance the washing of \rebuttal{the} LAA\rebuttal{,} and whether this habit will significantly impact \rebuttal{the} risk of thrombosis in the medium and long term. 

\section{Methods}
\label{sec:methods}
\begin{figure*}[t]
\centerline{\includegraphics[width=0.95\linewidth]{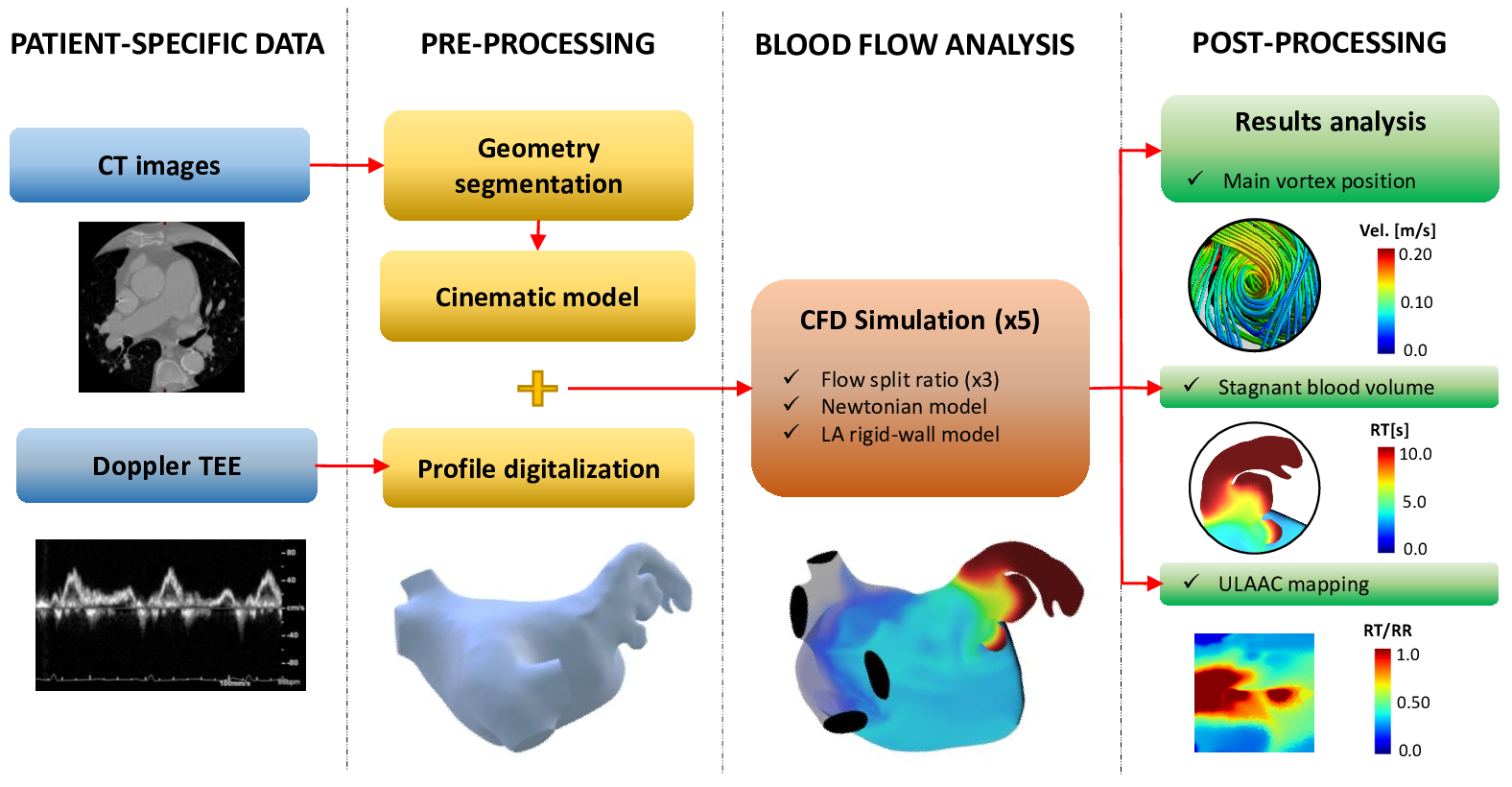}}
\caption{Workflow of the study. \textit{Patient-specific data:} CT images and Doppler TEE. \textit{Data pre-processing:} cardiac CT images were segmented to obtain the geometric atrial models. CFD-ready volumetric meshes were generated from each patient-specific geometry. Each Doppler curve was digitalized and adjusted to account for the different flow split ratios. \textit{Flow analysis:} Five different simulations were performed for each patient-specific geometry. \textit{Post-processing:} Three different analyses were performed: main vortex position study, SBV calculation, and ULAAC mapping.}
\label{fig:workflow}
\end{figure*}

In this \rebuttal{study}, we \rebuttal{analyzed} the LA geometries of nine patients. Our analysis \rebuttal{involved} several procedures, including \rebuttal{computed} tomography (CT), \rebuttal{image} segmentation, extraction of boundary conditions from Doppler transesophageal echocardiography (TEE), implementation of wall motion, CFD simulation, and \rebuttal{post-processing} of the results. The workflow \rebuttal{of this study} is summarized in Figure \ref{fig:workflow}.

\subsection{Patient-specific medical data acquisition}
\label{subsec:medical_data}

Table \ref{tab:clinic-data} presents the clinical data of the nine patients in the cohort, displaying both population data (age, sex, \rebuttal{clinical} history, \rebuttal{and} type of AF) and patient-specific mechanistic values (LA/LAA volume \rebuttal{and} HR). The Cardiology Service of the University Hospital of Badajoz (Spain) provided \rebuttal{the} CT and Doppler TEE images \rebuttal{for} each patient. \rebuttal{The} Hospital Ethics Committee approved the study \rebuttal{according to the Declaration of Helsinki guidelines} and all participants \rebuttal{provided written} informed consent.

\begin{table*}[ht]
\centering
\caption{Patient cohort and clinical data}

\begin{tabular}{lccccccccc}
 & \textbf{Case 1} & \textbf{Case 2} & \textbf{Case 3} & \textbf{Case 4} & \textbf{Case 5} & \textbf{Case 6} & \textbf{Case 7} & \textbf{Case 8} & \textbf{Case 9}\\ \hline
\textbf{Age [years]} & 89 & 67 & 54 & 72 & 83 & 71 & 88 & 79 & 86\\ 
\textbf{Sex [M/F]} & F & M & M & M & M & M & M & M & M \\ 
\textbf{Clinic history} & HT/DLP & HT & HT & HT & HT & HT/DM & HT/DLP & HT/DLP/DM & HT/DM \\ 
\textbf{AF type} & PE & PE & PE & PA & PA & PA & PA & PE & PE \\ 
\textbf{LA volume [mL]} & 114-138 & 180-205 & 228-252 & 68-92 & 153-175 & 245-269 & 88-107 & 220-228 & 137-157 \\ 
\textbf{HR [bpm]} & 50 & 95 & 86 & 60 & 75 & 125 & 45 & 82 & 52\\ 
\textbf{LAA volume [mL]} & 10 & 6 & 16 & 4 & 19 & 20 & 8 & 14 & 5 \\ \hline
\end{tabular}
HT\rebuttal{,} hypertension; DM\rebuttal{,} diabetes mellitus; DLP\rebuttal{,} cholesterol;  HR\rebuttal{,} heart rate; PE\rebuttal{,} Permanent; PA\rebuttal{,} Paroxysmal\\ 
\label{tab:clinic-data}
\end{table*}

CT images were \rebuttal{obtained using} a LightSpeed VCT \rebuttal{(}General Electric Medical \rebuttal{System,} Milwaukee, WI, USA). This scanner has 64 detectors and is equipped with snapshot segment\rebuttal{-}mode technology, which enables precise cardiac imaging. The scanning parameters were \rebuttal{as follows}: slice thickness\rebuttal{,} 0.625 mm; rotation time\rebuttal{,} 0.4 s; tube voltage\rebuttal{,} 120 kV; collimation\rebuttal{,} 0.625 mm; and pitch\rebuttal{,} 0.18, 0.20, 0.23, or 0.26, determined by software based on each patient's heart rate.

\begin{figure*}[t]
\centerline{\includegraphics[width=0.9\linewidth]{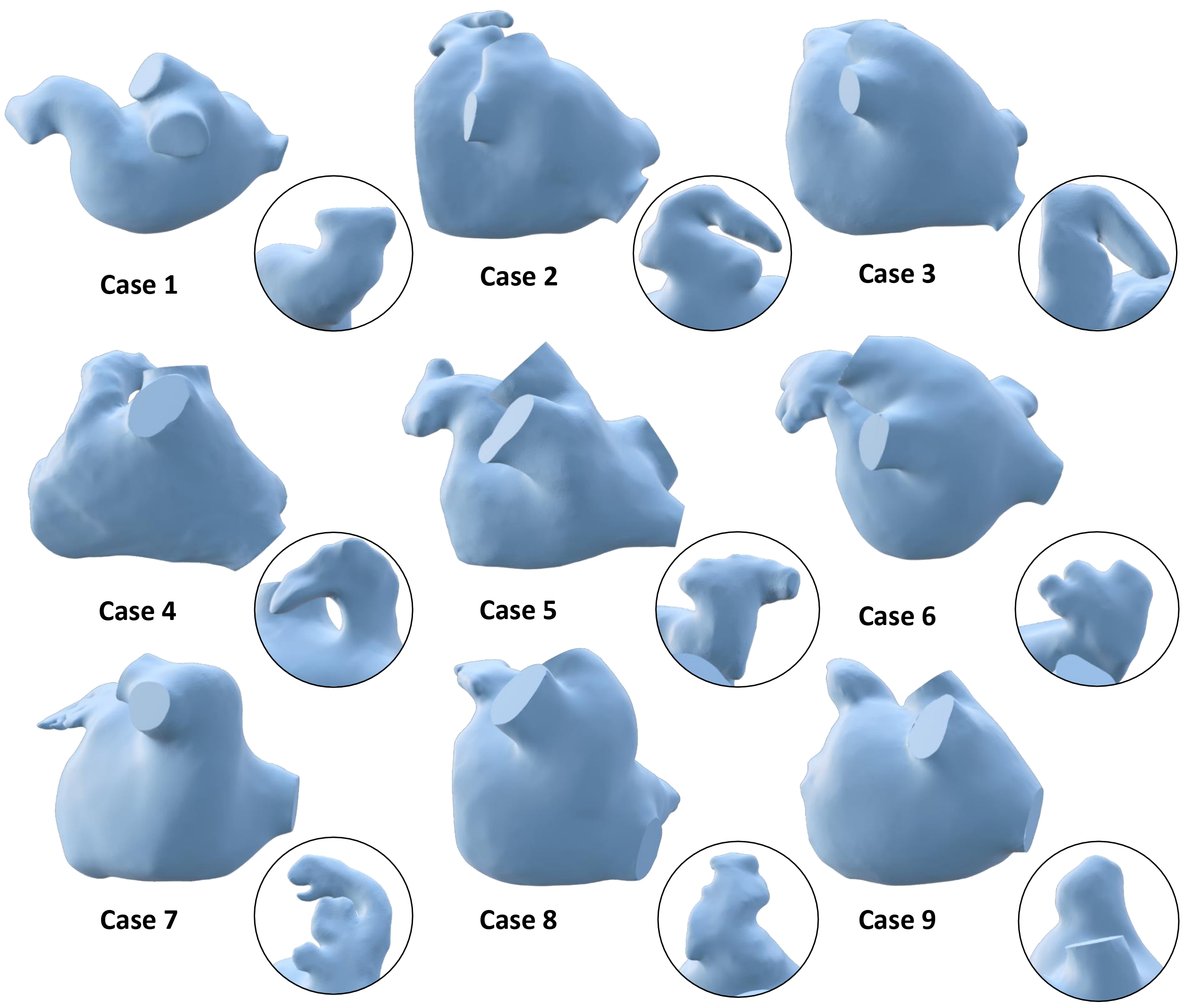}}
\caption{Patient-specific left atrial structures in 0\%RR.  The structure of the LA is represented in the posterior view, and the structure of the LAA is represented in the anterior view.}
\label{fig:geoms}
\end{figure*}

A 3D model generation procedure was implemented using semi-automatic segmentation \cite{duenas2021comprehensive}. The procedure was validated by a radiologist and an interventionist cardiologist. \rebuttal{The surface} meshes obtained from the segmentation were cleaned and prepared using Blender \cite{Blender2023}. The resulting geometries are shown in Figure \ref{fig:geoms}. The atrial surfaces are represented in 0\%RR, where RR indicates the patient-specific duration of the cardiac cycle between two consecutive electrocardiographic R waves, and 0\%RR indicates the end-diastole of the electrocardiographic ventricular.

The velocities in the PV and the MV were measured using two Philips cardiovascular ultrasound systems (EPIC CVx and Affiniti CVx). Each of the five images obtained for each patient was digitalized using an in-house code written in Matlab (Mathworks, Inc., Natick, MA, USA). The Doppler waveforms were conveniently synchronized with the CT images, taking as a reference 0\%RR. The inlet boundary conditions were established by imposing \rebuttal{a} flow through each PV, \rebuttal{as} calculated from the PV Doppler velocities. The MV Doppler velocity profile \rebuttal{is} used to validate the simulation. 

\subsection{Parametrization of the LA wall motion}
\label{subsec:wall_motion}

Patient-specific geometries and parametric \rebuttal{displacements} were used to compensate for the lack of information on the LA displacement field \rebuttal{because} the data \rebuttal{were} scanned for each patient \rebuttal{at} a single time instant. Following the \rebuttal{works} of \cite{corti2022impact,zingaro2021hemodynamics,zingaro2021geometric}, we prescribed a smooth movement \rebuttal{toward} the atrial center of mass. The code algorithm \rebuttal{for} the implementation described in this subsection can be found \rebuttal{in} \textit{Algorithm 1} in the Supplementary Material. Let $\Omega_t$ be the fluid domain at a specific time $t \in (0,T)$ and $\partial \Omega_t$ its boundary, where $T$ is the cardiac period. Then we assume that the boundary velocity $\boldsymbol{g}^{ALE}: \partial \Omega_t \times (0,T) \rightarrow \mathbb{R}^3$ can be expressed as a separation of variables:

\begin{equation}
    \boldsymbol{g}^{ALE}(\boldsymbol{\hat{x}},t) = \boldsymbol{F}^{ALE}(\boldsymbol{\hat{x}}) \cdot h^{ALE}(t) \quad \text{ on } \partial \Omega_t \times (0,T).
    \label{eq:g_ALE}
\end{equation}

We will now explain how to construct the functions $h^{ALE}(t)$ and $\boldsymbol{F}^{ALE}(\boldsymbol{\hat{x}})$:

\begin{equation}
\frac{dV_{LA}}{dt} = \frac{d}{dt} \int_{\Omega_t} d\boldsymbol{x} \overset{\textsf{(RTT)}}{=} \int_{\partial \Omega_t} \boldsymbol{g}^{ALE} (\boldsymbol{\hat{x}},t) \cdot \boldsymbol{n}  d\sigma,
\label{eq:rtt}
\end{equation} 

where $V_{LA}(t)$ is the volume of the LA and RTT, the Reynolds Transport Theorem (RTT) \cite{kundu2015fluid}. The time-dependent function for $h^{ALE}(t)$ can be obtained by combining Equations \ref{eq:g_ALE} and \ref{eq:rtt} as:

\begin{equation}
    h^{ALE}(t) = \left( \int_{\partial \Omega_t} \boldsymbol{F}^{ALE} (\boldsymbol{\hat{x}}) \cdot \boldsymbol{n} (\boldsymbol{\hat{x}}, t) d \sigma \right)^{-1} \frac{dV_{LA} (t)}{dt}.
\end{equation}

\begin{figure*}[p] 
    \centering
   \begin{subfigure}[t]{0.87\textwidth}
        \centering
        \includegraphics[width=\linewidth]{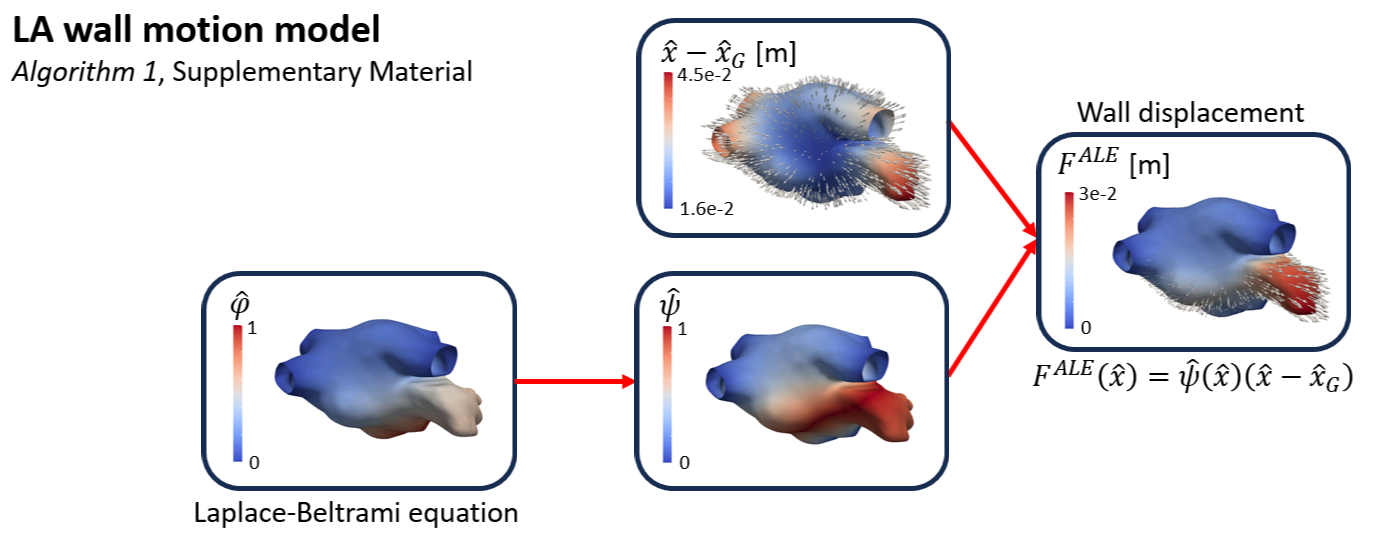}
        \caption{}
          \label{subfig:method_wall_motion}
    \end{subfigure}
    \begin{subfigure}[t]{0.87\textwidth}
        \centering
        \includegraphics[width=\linewidth]{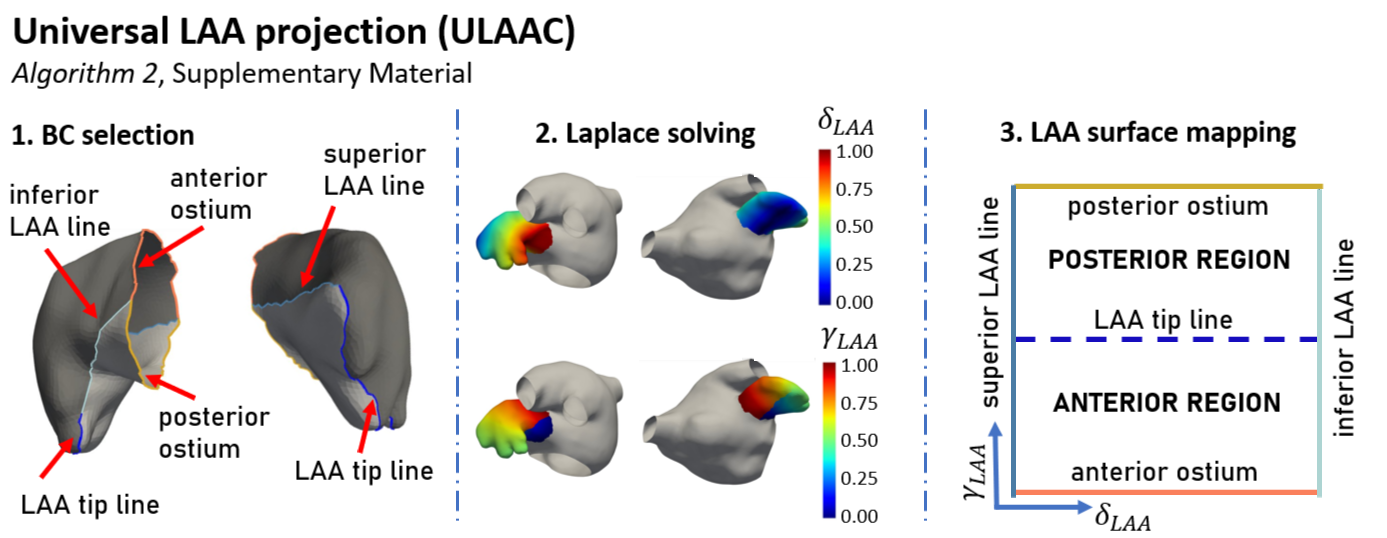}
        \caption{}
          \label{subfig:method_ulaac}
    \end{subfigure}    
    \begin{subfigure}[t]{0.87\textwidth}
        \centering
        \includegraphics[width=\linewidth]{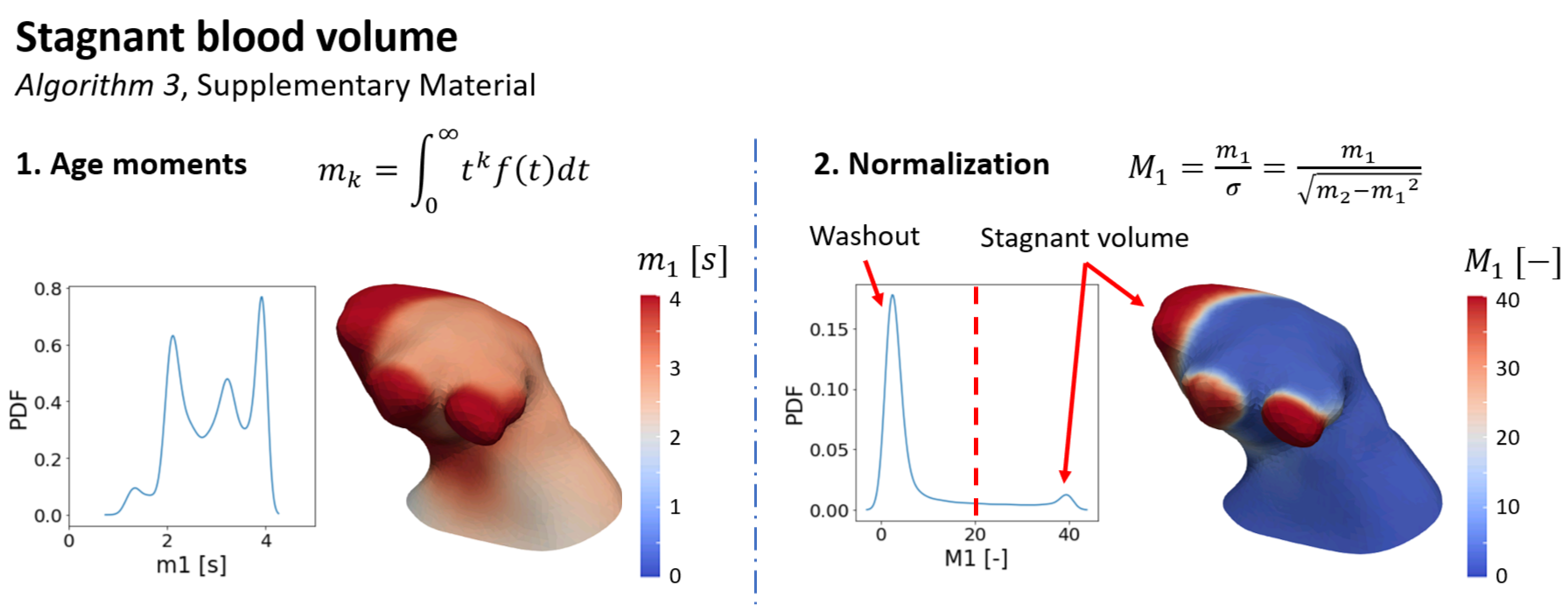}
        \caption{}
          \label{subfig:method_stagnant_volume}
    \end{subfigure} 
\caption{Outline of the methods used throughout this manuscript. (a) LA wall motion model \cite{corti2022impact,zingaro2021hemodynamics,zingaro2021geometric} to normalize fibrillated motion between patients. (b) \rebuttal{ULAAC} system \cite{duenas2024reduced} to facilitate data visualization and comparison between different patients. (c) Calculation of the SBV \cite{duenas2021comprehensive, duenas2022morphing} to delineate the prothrombotic LAA region.} 
\label{fig:methods}
\end{figure*}

On the other hand, the space-dependent component of $\boldsymbol{F}^{ALE} (\boldsymbol{\hat{x}})$ controls the motion of the entire chamber and is defined as:

\begin{equation}
    \boldsymbol{F}^{ALE}(\boldsymbol{\hat{x}}) = \hat{\Psi} (\boldsymbol{\hat{x}}) \left( \boldsymbol{\hat{x}} - \boldsymbol{\hat{x}}_G \right),
\end{equation}

where $\boldsymbol{\hat{x}}_G$ is a vector field that points toward the center of mass of the atrium. The function $\hat{\Psi}$ is computed as a normalized product:

\begin{equation}
    \hat{\Psi} (\boldsymbol{\hat{x}}) =  \frac{\hat{\varphi} (\boldsymbol{\hat{x}}) \left( 1 - \hat{\varphi} (\boldsymbol{\hat{x}}) \right) }{\max_{ \boldsymbol{\hat{x}} \in \partial \hat{\Omega} } \{ \hat{\varphi} (\boldsymbol{\hat{x}}) \left( 1 - \hat{\varphi}(\boldsymbol{\hat{x}}) \right) \}},
\end{equation}

where $\hat{\varphi}$ is the solution of a \rebuttal{Laplace--Beltrami} problem \cite{bonito2020finite}. This problem is defined as follows: 

\begin{equation}
        \begin{cases}
      - \Delta \hat{\varphi} = 0 \quad \text{ in } \partial \hat{\Omega},  \\
      \hat{\varphi} = 0 \quad \quad \quad \text{ on } \bigcup^{4}_{j=1} \hat{\Gamma}^{PV_j}, \\
      \hat{\varphi} = 1 \quad \quad \quad \text{ on } \hat{\Gamma}^{MV},
    \end{cases} 
\end{equation}

where $\hat{\Gamma}^{PV_j}$ is the $j^{th}$ PV inlet section (with $j = 1, ..., 4$) and $\hat{\Gamma}^{MV}$ the MV outlet section. Using the given definitions, we have derived a smooth function that maps $\partial \hat{\Omega}$ to the interval $[0,1]$. This function is zero in the inlet and outlet sections of the computational domain (PV and MV\rebuttal{, respectively}). The steps described to calculate $\boldsymbol{F}^{ALE}(\boldsymbol{\hat{x}})$ are shown in Figure \ref{subfig:method_wall_motion}. Following the steps mentioned above, we successfully calculated the boundary velocity $\boldsymbol{g}^{ALE}(\boldsymbol{x},t)$ for each patient-specific flow profile obtained from Doppler velocities, thus ensuring the consistency of the mass flux through the domain.

\subsection{Patient-specific numerical model}
\label{subsec:patient_specific_model}

ANSYS Meshing (ANSYS, Inc.) was used to generate tetrahedral meshes for each patient-specific geometry at 0\% RR. We performed a mesh convergence study to confirm that the resolution of the meshes did not significantly affect the simulation results. The total number of cells for the selected meshes ranged from 1.5M to 2M, showing negligible velocity differences between the selected mesh and a more refined one for our particular geometries. This is important since spatiotemporal resolution can influence predicted hemodynamics and stresses \cite{Khalili2024}.

ANSYS Fluent software (ANSYS, Inc.) was used to perform the simulations. Both the advection and transient schemes were second-order accurate, and we set a convergence criterion of $10^{-4}$ for the residuals. We chose the \rebuttal{coupled} scheme as the pressure-velocity coupling algorithm.

The dynamic mesh capabilities of ANSYS were applied to prescribe the wall motion generated by the kinematic model described in \rebuttal{Section} \ref{subsec:wall_motion}. A smoothing method employing a node diffusion algorithm with \rebuttal{the} cell volume as \rebuttal{the} diffusion function \rebuttal{was used}. Atrial motion was severely reduced \rebuttal{because of} AF, so re-meshing techniques were unnecessary.

As mentioned in \rebuttal{Section} \ref{subsec:medical_data}, \rebuttal{the} Doppler velocity measurements in the PV \rebuttal{provide} patient-specific boundary conditions at the inlet, while \rebuttal{the} measurements in the MV allowed simulation results to be validated. The boundary conditions were established by imposing \rebuttal{an} inlet flow and a constant outlet pressure. \rebuttal{Inlet flows} obtained from Doppler measurements \rebuttal{were} adjusted to generate different flow split cases 59-41, 50-50, and 37-63 LPV-RPV\%.

The fluid density was assumed to be $\rho = \SI{1050}{kg/m^3}$ and the dynamic viscosity $\mu$ obeys Carreau's rheological model \cite{bird87}: 

\begin{equation} \label{eq:carreau}
\mu =\mu_{\infty}+\omega\left(\mu_0-\mu_{\infty}\right)\left[1+\left(\lambda\dot\gamma\right)^2\right]^\frac{n-1}{2},
\end{equation}
where $\dot\gamma$ is the shear rate, $\mu_{\infty}=\SI{0.00345}{\pascal\cdot\second}$ and $\mu_0=\SI{0.056}{\pascal\cdot\second}$ are the infinite and zero shear rate viscosities, $\lambda=\SI{3.31}{\second}$ is the characteristic time, and $n=0.3568$ is the power-law index. All these values \rebuttal{have been} previously validated \cite{CK91}. The non-Newtonian character is provided by the dimensionless number $\omega$ ($\omega=0$ and $1$ would correspond to a Newtonian and Carreau model, respectively).

The continuity and Navier-Stokes equations (\rebuttal{Equations} \ref{eq:continuity} and \ref{eq:momentum}, respectively) \rebuttal{are} solved for each model\rebuttal{.}

\begin{equation} \label{eq:continuity}
\nabla \cdot \textbf{v} = 0,
\end{equation}

\begin{equation} \label{eq:momentum}
\frac{\partial{\textbf{v}}}{\partial{t}} + \textbf{v} \cdot \nabla \textbf{v} = - \frac{ \nabla p}{\rho} + \frac{\mu}{\rho} {\nabla}^2 \textbf{v},
\end{equation}\\

where $\textbf{v}$ is the velocity vector and $p$ is the pressure.

A time convergence analysis was performed to verify the time convergence \rebuttal{of the solution}. The flow velocity and hemodynamic metrics \rebuttal{were monitored} at several random points within the domain, \rebuttal{and} a quasiperiodic regime \rebuttal{was reached} after four cycles. Therefore, the \rebuttal{subsequent} analysis \rebuttal{was} based on the results \rebuttal{of} the fourth cycle.  
Finally, the simulation time step $\Delta t=\SI{0.001}{\second}$ was halved, and we did not detect an appreciable variation in the results confirming the use of the mentioned time step.

\subsection{Universal left atrial appendage coordinate projection}
\label{subsec:ulaac}

We used the \rebuttal{universal left atrial appendage coordinate} (ULAAC) system \cite{duenas2024reduced} to facilitate data visualization and compare \rebuttal{the} simulation results between different anatomical models. \rebuttal{This} approach involves \rebuttal{the} mapping \rebuttal{of} various LAA surfaces onto \rebuttal{a} unit square. This was achieved by solving the Laplace equation on each LAA surface for two coordinates, the infero-posterior position $\delta_{LAA}$ and the antero-posterior position $\gamma_{LAA}$. Boundary conditions were defined at some anatomical landmarks, \rebuttal{as} shown in Figure \ref{subfig:method_ulaac}: the anterior and posterior ostium, the LAA tip line, and the superior and inferior LAA lines.

The surface mesh \rebuttal{constituting} the anterior LAA region \rebuttal{was} defined by a connected path that \rebuttal{included} the superior LAA line, the tip line, the inferior LAA line, and the anterior ostium. Similarly, the surface mesh that constitutes the posterior LAA region is defined by the connected path formed by the superior LAA line, the tip line, the inferior LAA line, and the posterior ostium.

Finally, to ensure a bijective correspondence, the anteroposterior coordinate $\gamma_{LAA}$ obtained by solving the Laplace equation was rescaled by a factor of 2 and unwrapped around 0.5. This transformation was \rebuttal{performed} to ensure that the \rebuttal{coordinates range} from 0 to 0.5\rebuttal{,} in the anterior LAA region (from the anterior ostium to the tip) and from 0.5 to 1\rebuttal{,} in the posterior LAA region (from the tip to the posterior ostium).  

The \rebuttal{Laplace--Beltrami} equations were solved using Python LaPy \cite{reuter2006laplace, wachinger2015brainprint}. The complete procedure is described in more detail by Dueñas \textit{et al.} \cite{duenas2024reduced}, and the code algorithm of the method described in this subsection can be found \rebuttal{in} \textit{Algorithm 2} in the Supplementary Material.

\subsection{Hemodynamic indices}
\label{subsec:indices}

\rebuttal{Similar} to our previous work \cite{balzotti2023reduced, duenas2021comprehensive,duenas2022morphing,duenas2024reduced}, we analyzed the moments of blood age \cite{Sierra2017} to determine \rebuttal{the} stagnant regions. To this end, we define the $m_k$ age moment as
\begin{equation}\label{eq:moment_def}
  m_k = \int_{0}^{\infty}{t^k f(t)\, dt},
\end{equation}
where $f(t)$ is the blood age distribution and $t$ is simulation time.
The normalized first moment ($M_1$) of the blood age distribution was computed,
  \begin{equation}
    M_1 = \frac{m_1}{\sigma},
  \end{equation}
where $m_1$ is the first moment of the distribution function (that is, the blood RT), and $\sigma$ its standard deviation:

\begin{equation}
    \sigma = \sqrt{m_2 - m_1^2}.
\end{equation}

$M_1$ presents a bimodal distribution within the LAA \cite{duenas2021comprehensive}, which allows the automatic selection of a patient-specific threshold to delineate the SBV. The minimum probability valley between the two modes was selected as the threshold (Figure \ref{subfig:method_stagnant_volume}).

The moment equations of the age distribution were integrated into the solver \rebuttal{using} Fluent \rebuttal{user-defined functions}. The complete procedure \rebuttal{is} described in detail \rebuttal{in} our previous work \cite{duenas2021comprehensive, Sierra2017}, and the code algorithm for the calculation of age moments can be found \rebuttal{in} \textit{Algorithm 3} in the Supplementary Material.

Some wall shear-based indicators, such as the time-averaged wall shear stress (TAWSS) and oscillatory shear index (OSI), were also calculated and compared during \rebuttal{the} blood flow analysis. \rebuttal{The} TAWSS is commonly used as \rebuttal{a} hemodynamic indicator \rebuttal{because} high values are associated with endothelial cell damage\rebuttal{, which} activates the extrinsic coagulation cascade \cite{rojas-gonzalez2023}. \rebuttal{The} OSI is a dimensionless indicator that captures \rebuttal{oscillations in the wall-shear direction} between 0 and 0.5. \rebuttal{These} are defined as follows:

\begin{equation}
\text{TAWSS} = \frac{1}{T} \int_0^T \lvert \text{WSS} \rvert dt,
\end{equation}

\begin{equation}
\text{OSI} = \frac{1}{2} \left(1 - \frac{\lvert \int_0^T  \text{WSS } dt \rvert }{\int_0^T \lvert \text{WSS} \rvert dt} \right).
\end{equation}

\section{Results}\label{sec:results}

45 atrial simulations were launched, 5 for each of the 9 patient-specific geometries (Figure \ref{fig:workflow}): 
\begin{itemize}
    \item \rebuttal{Three} simulations considering one of the following flow split ratios: 59-41, 50-50, and \rebuttal{37-63\%LPV-RPV} to compare the impact of extreme flow split ratios. \rebuttal{These studies} considered the same prescribed wall motion and a non-Newtonian model.   
    \item One simulation \rebuttal{considered} rigid walls to analyze the impact of wall motion. \rebuttal{These simulations} considered a non-Newtonian model \rebuttal{with} a flow split ratio of \rebuttal{50-50\%LPV-RPV}.
    \item  One simulation \rebuttal{used} a Newtonian model to compare the \rebuttal{effect} of \rebuttal{the} non-Newtonian model. This simulation considered a prescribed wall motion and a flow split ratio of \rebuttal{50-50\%LPV-RPV}. 
\end{itemize}

\subsection{Representative atrial flow patterns}

LA flow is generally characterized by \rebuttal{the} circulatory flow generated by the LPV, while \rebuttal{the} RPV flows directly into the MV \cite{Vedula2015}. These atrial flow patterns \rebuttal{are} visualized for each patient by calculating the blood flow streamlines. As one of the goals of this \rebuttal{study} was to clarify how atrial flow patterns affect hemodynamics in \rebuttal{the} LAA, we classified atrial flow behavior according to the position of the center of the circulatory flow of \rebuttal{the} LPV that governs LA flow: right after \rebuttal{the} LPV (type A), near the LAA ostium (type B1)\rebuttal{,} or near the LA roof (type B2). For simplicity, we establish as a reference a flow split ratio of \rebuttal{50-50\%LPV-RPV}, a prescribed wall motion, and a Newtonian model. 

\begin{figure*}[p] 
    \centering
   \begin{subfigure}[t]{0.87\textwidth}
        \centering
        \includegraphics[width=\linewidth]{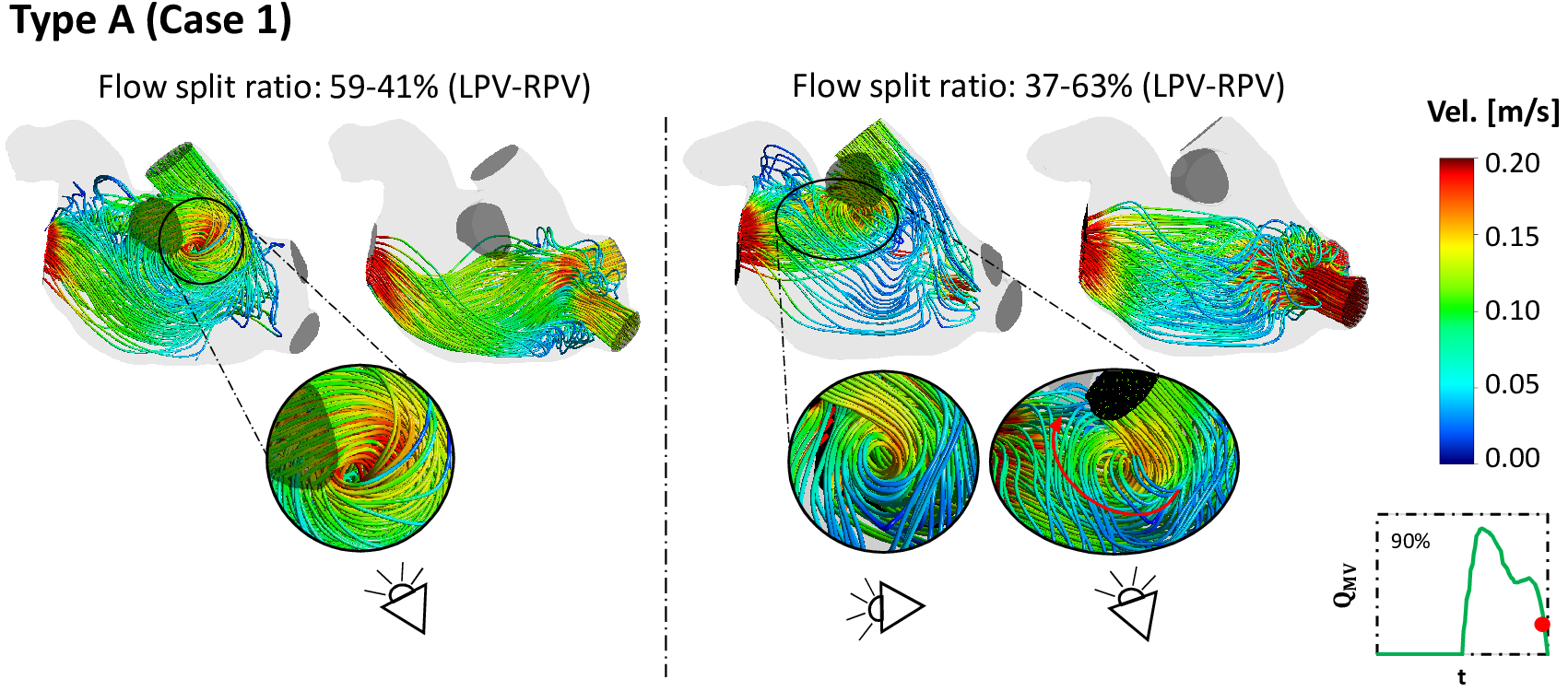}
        \caption{}
          \label{subfig:streamlines_A}
    \end{subfigure}
    \begin{subfigure}[t]{0.87\textwidth}
        \centering
        \includegraphics[width=\linewidth]{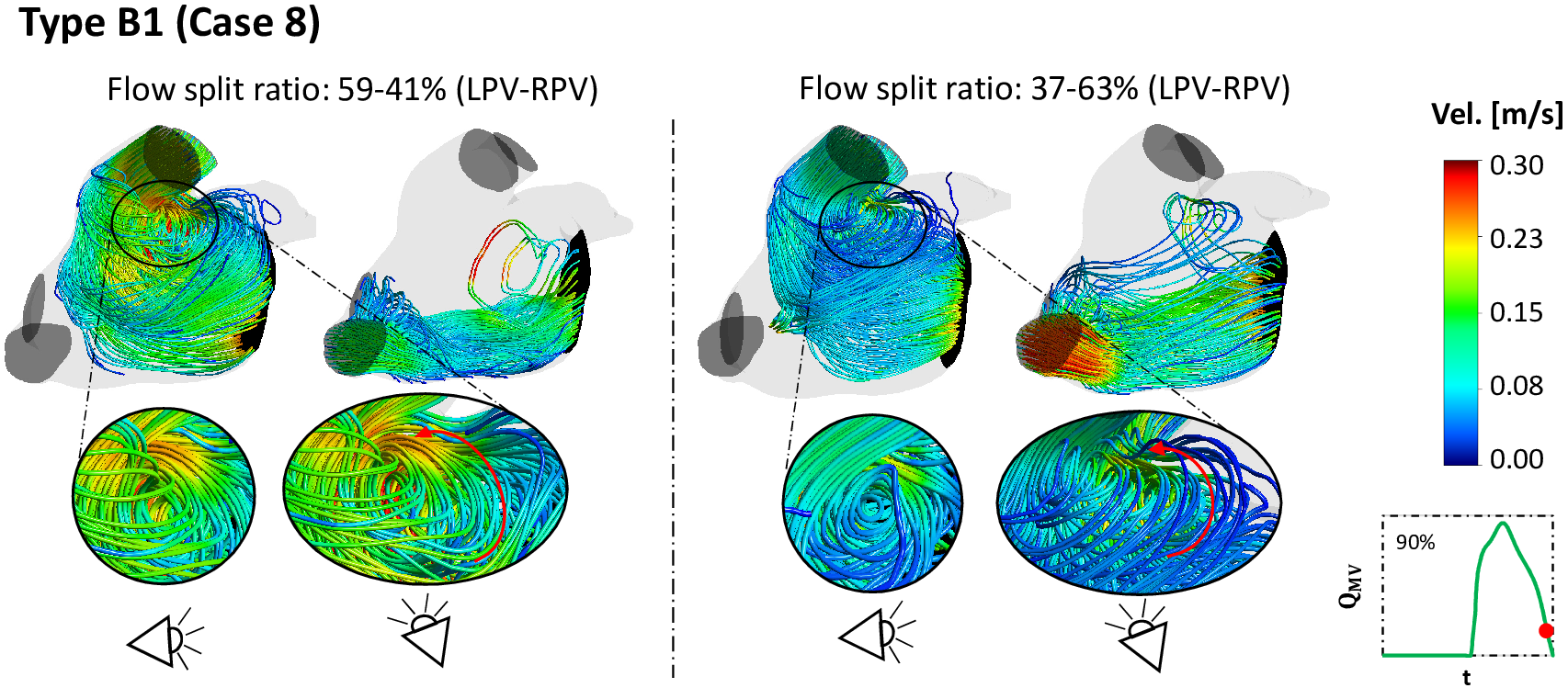}
        \caption{}
          \label{subfig:streamlines_B1}
    \end{subfigure}    
    \begin{subfigure}[t]{0.87\textwidth}
        \centering
        \includegraphics[width=\linewidth]{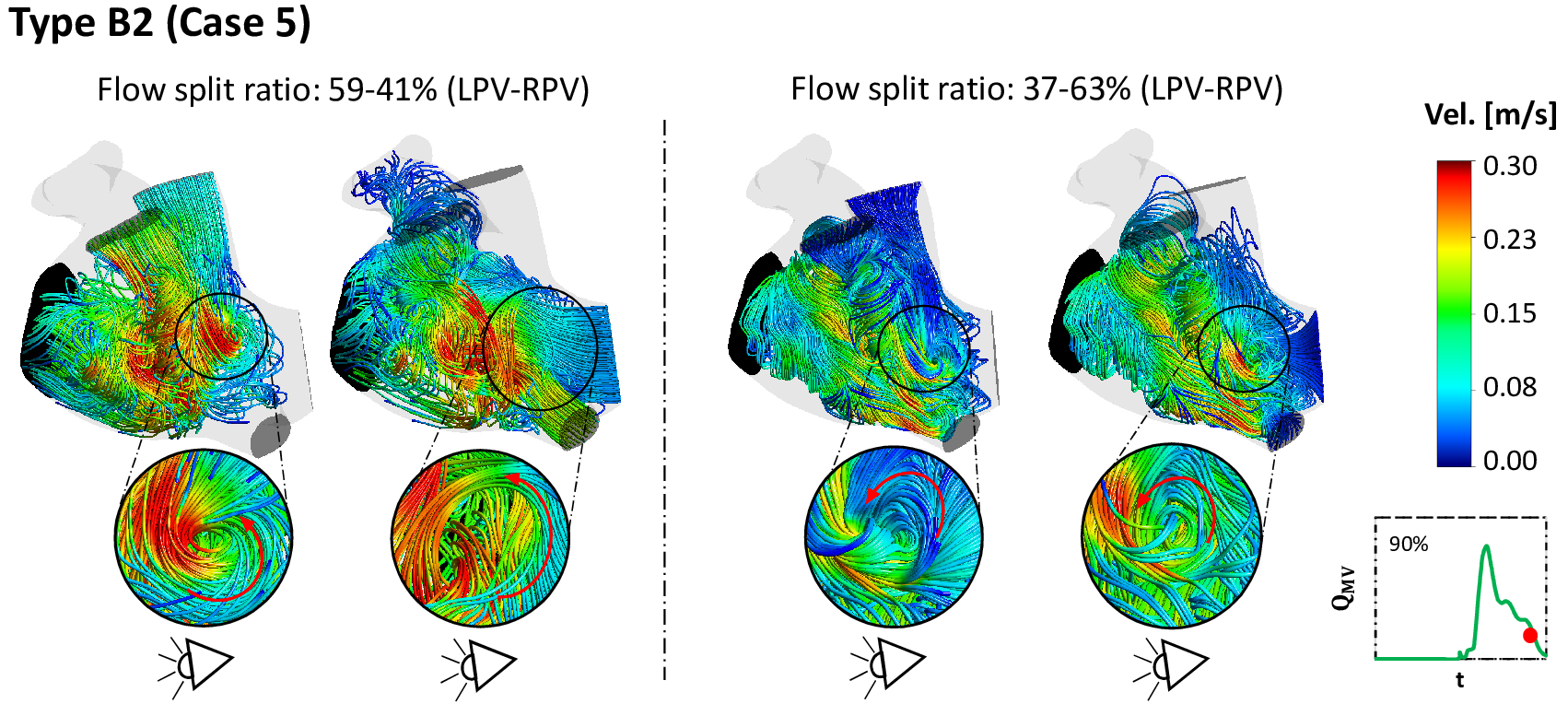}
        \caption{}
          \label{subfig:streamlines_B2}
    \end{subfigure} 
\caption{Streamlines of blood flow at 90\%RR for the three representative cases: (a) Type A, (b) Type B1, and (c) Type B2. Two different flow split ratios are depicted for each representative case. The right and left renderings for each flow split ratio illustrate the streamlines coming from LPV and RPV, respectively. LA was rotated to facilitate flow visualization and zooms and view angles were added. Arrows indicate the direction of rotation.}
\label{fig:velocity_streamlines}
\end{figure*}

Following this classification, each row in Figure \ref{fig:velocity_streamlines} shows a representative case of one of these flow patterns, which facilitates a better understanding of the interaction between the LPV-RPV flow, main atrial circulatory flow, and LAA washout. We selected to display the instant of the cycle 90\% RR since the main circulatory flow \rebuttal{was} fully developed. For each flow type, two different flow split ratios are shown for comparison: 59-41 and 37-63\% LPV-RPV. \rebuttal{Similarly}, each flow split column is divided into two subcolumns: the left subcolumn displays the streamlines from the LPV, while the right subcolumn displays the streamlines \rebuttal{originating} from the RPV.

Regarding type A atrium (Figure \ref{subfig:streamlines_A}), when the flow split is 59-41\%, the main atrial circulatory flow \rebuttal{originated immediately} after the LPV. In contrast\rebuttal{, the} washout flow \rebuttal{could not} be appreciated within the LAA. However, when the flow split \rebuttal{was} changed to 37-63\%, the main circulatory flow \rebuttal{was} carried downstream near the ostium, facilitating the generation of a secondary circulatory flow within the LAA. Although the RPV flow does not play a direct role in LAA washing (\rebuttal{because} its flow is directed toward the MV), it can push the main circulatory flow toward the ostium, facilitating washing of the LAA. 

\rebuttal{In the} type B1 atrium (Figure \ref{subfig:streamlines_B1}), the main circulatory flow \rebuttal{was} located near the ostium when the flow split \rebuttal{was} 59-41\%, allowing the LAA to be washed. When the flow split \rebuttal{was} changed to 37-63\%, the RPV flow slightly \rebuttal{pushed} the position of the main circulatory flow. However, \rebuttal{a} reduction in LPV flow reduces LAA washing, which \rebuttal{is} governed by a secondary flow induced by the primary circulatory flow of the LPV. \rebuttal{Similar} to the previous case, the RPV flow \rebuttal{did} not directly influence the LAA flow, which \rebuttal{was} entirely washed \rebuttal{off by} LPV secondary flows. 

\rebuttal{Finally}, for \rebuttal{a} type B2 atrium (Figure \ref{subfig:streamlines_B2}), the orientation of the LPV results in the formation of the main atrial circulatory flow near the atrial roof and far from the ostium. This results in a more chaotic flow: for a flow split of 59-41\%, the RPV flow passes around the main circulatory flow, while for the 37-63\% case, the RPV flow collides and mixes with the main circulatory flow. In the first case, the washing of the LAA is governed by the RPV flow \rebuttal{surrounding} the main circulatory flow, \rebuttal{whereas} in the second case, no secondary flow is induced in the LAA.

\subsection{Vortex structure analysis}
\begin{figure*}[p] 
    \centering
   \begin{subfigure}[t]{0.9\textwidth}
        \centering
        \includegraphics[width=\linewidth]{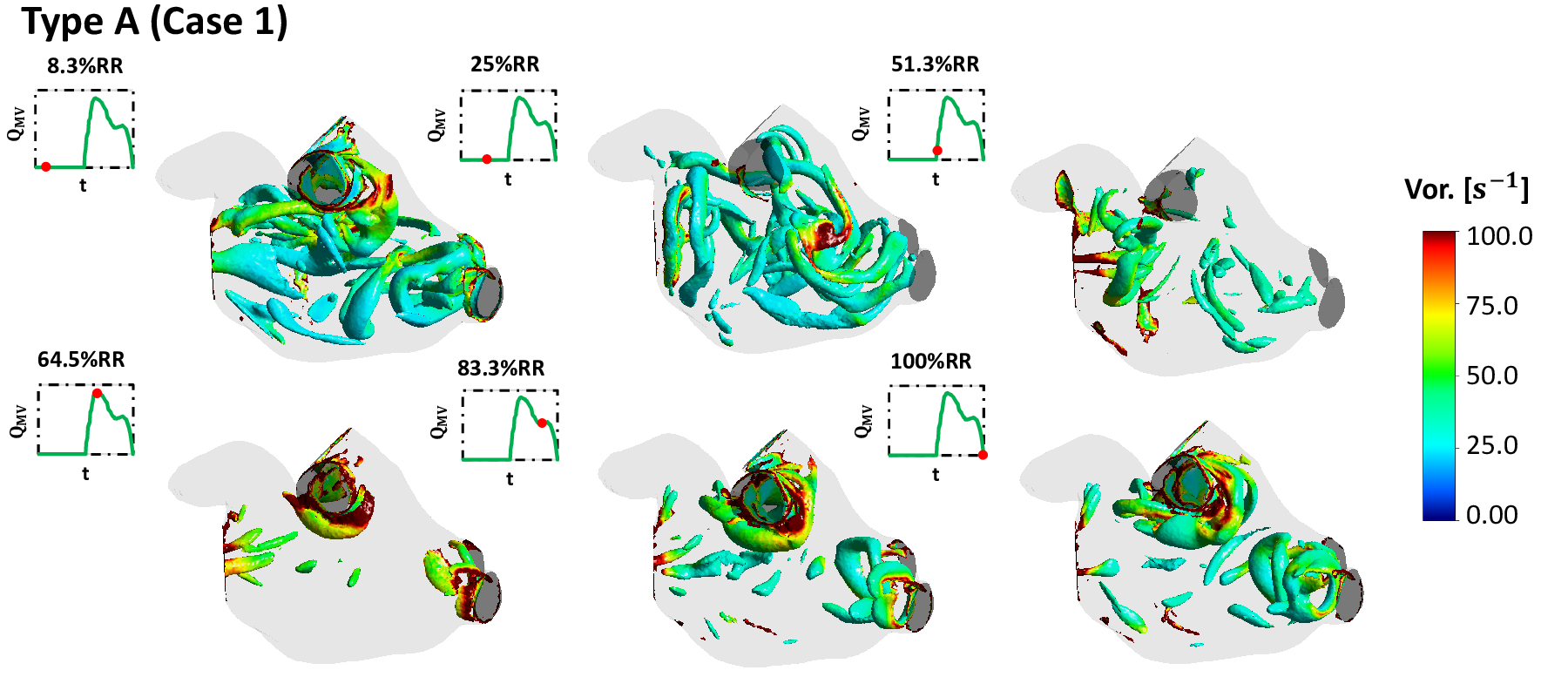}
        \caption{}
          \label{subfig:vor_vortex_A}
    \end{subfigure}
    \begin{subfigure}[t]{0.9\textwidth}
        \centering
        \includegraphics[width=\linewidth]{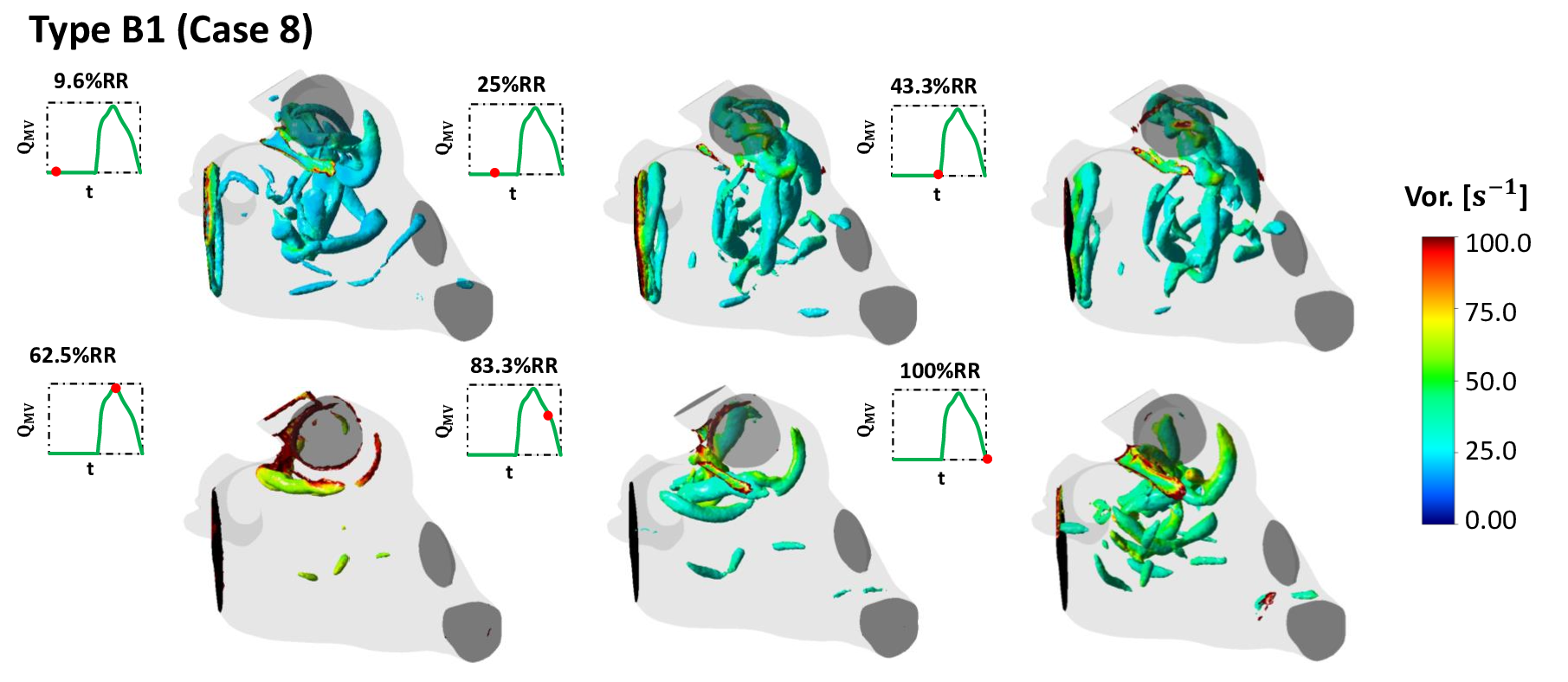}
        \caption{}
          \label{subfig:vor_vortex_B1}
    \end{subfigure}    
    \begin{subfigure}[t]{0.9\textwidth}
        \centering
        \includegraphics[width=\linewidth]{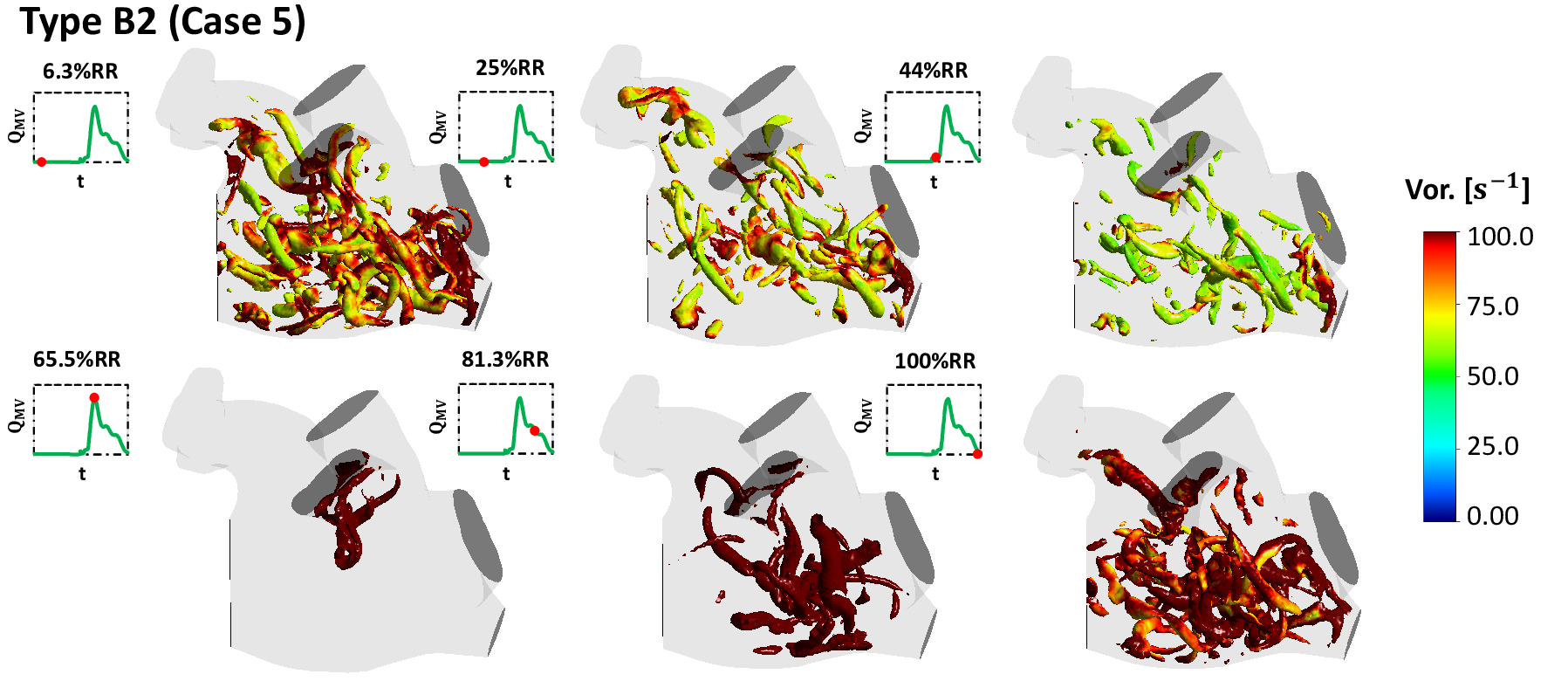}
        \caption{}
          \label{subfig:vor_vortex_B2}
    \end{subfigure} 
\caption{Vortex structures during a cardiac cycle for the three representative cases: (a) Type A, (b) Type B1, and (c) Type B2. The structures were generated following the $q_2$ criterion ($q2 = 0.2$) and colored by the magnitude of the vorticity. }
\label{fig:vor_vortex_evo}
\end{figure*}

In the previous section, we observed flow patterns within the LA and classified anatomies \rebuttal{into} three types \rebuttal{by} analyzing the flow behavior by visualizing the streamlines. We now proceed to confirm these considerations by vortex structure analysis, \rebuttal{using} the same simulation as reference (\rebuttal{50-50\%LPV-RPV}, a prescribed wall motion\rebuttal{,} and a Newtonian model).
Figure \ref{fig:vor_vortex_evo} shows the characteristic vortex structures at different cardiac cycle instants for each representative position of the atrial vortex, where the rows represent types A, B1, and B2, \rebuttal{and the} initial three instants denote systole, followed by three instants representing diastole. \rebuttal{Specifically}, the third instant marks the opening of \rebuttal{the} MV, the fourth corresponds to the peak flow phase through the PV, the fifth to a fully developed main circulatory flow, and the sixth to the closure of the MV. In the temporal progression of vortex structures within a type A atrium (Figure \ref{subfig:vor_vortex_A}), the first two instants (8.3\%RR and 25\%RR) \rebuttal{corresponded} to ventricular systole. During this part of the cycle, the LA is filled with \rebuttal{the} PV blood\rebuttal{, whereas} the MV is closed. As the flow \rebuttal{was} not conducted directly toward the MV, many small vortex structures \rebuttal{were} formed in the atrium. After opening the MV, formation of the main circulatory LPV flow \rebuttal{began}. The main circulatory flow structure for type A atriums is located \rebuttal{immediately} after the outlet of the LPV and \rebuttal{was} already fully developed \rebuttal{at} 83.3\% RR and 100\% RR. Most vortices \rebuttal{were} situated around this main circulatory flow, although some minor vortex structures \rebuttal{were also observed} within the LAA near the ostium and at the outlet of the RPV.

Regarding the vortex structures \rebuttal{of} type B1 (Figure \ref{subfig:vor_vortex_B1}), the vortex patterns \rebuttal{were} similar to those of type A except for the main vortex position, which \rebuttal{was} displaced downstream of the LPV \rebuttal{toward} the ostium. \rebuttal{Minor vortices} form within the LA during \rebuttal{the} ventricular systole when the MV is closed. In this case, \rebuttal{a few} vortical structures are formed in the RPV. Finally, \rebuttal{in} type B2 (Figure \ref{subfig:vor_vortex_B2}), a \rebuttal{significantly} different vortex pattern \rebuttal{occured} during ventricular diastole. Here, the RPV stream does not flow in an orderly manner straight to the MV but collides with the LPV stream, forming a more chaotic vortex pattern. This \rebuttal{was} reflected in the number and intensity of \rebuttal{vortices} within the LA for type B2.

\subsection{Left atrial appendage residence time}
\label{subsec:LAA_RT}

\begin{figure*}[t]
\centerline{\includegraphics[width=0.92\linewidth]{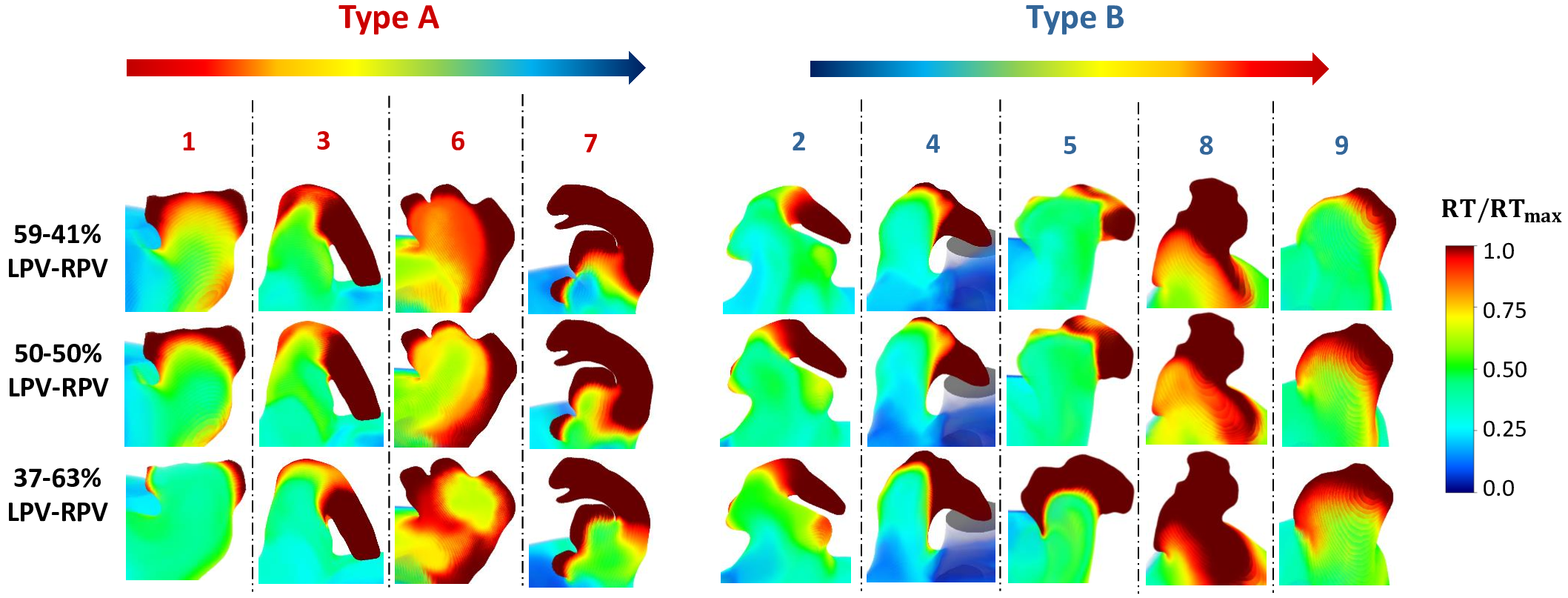}}
\caption{$\text{RT}/\text{RT}_{max}$ contours for the different flow split ratios (rows) and the nine patient-specific cases (columns). The cases are grouped into types A and B, respectively.}
\label{fig:LAA_RT_by_type}
\end{figure*}

To highlight the different flow behaviors within the LAA depending on the position of the main atrial vortex, the $\text{RT}/\text{RT}_{max}$ contours organized by the two main types (B1 and B2 are displayed together as type B for the sake of clarity) are shown in Figure \ref{fig:LAA_RT_by_type}. RT values were normalized with the $\text{RT}_{max}$ value for each patient. Type A is depicted on the left side of the figure, \rebuttal{whereas Type} B is shown on the right. \rebuttal{Subcolumns} represent different patients, \rebuttal{whereas} rows represent different flow split cases. 

As expected, for all cases, the regions with the highest $\text{RT}/\text{RT}_{max}$ are concentrated in the tip of the LAA, with the lower $\text{RT}/\text{RT}_{max}$ near the ostium. However, a different flow behavior can be observed within the LAA depending on the position of the main atrial vortex. For type A atriums, the $\text{RT}/\text{RT}_{max}$ within the LAA tends to decrease as the RPV flow fraction increases. Thus, the LAA washing increases with the LPV flow fraction. In contrast, case B presents the opposite trend: $\text{RT}/\text{RT}_{max}$ within the LAA tends to increase as the RPV flow fraction increases.

\begin{figure*}[t]
\centerline{\includegraphics[width=0.92\linewidth]{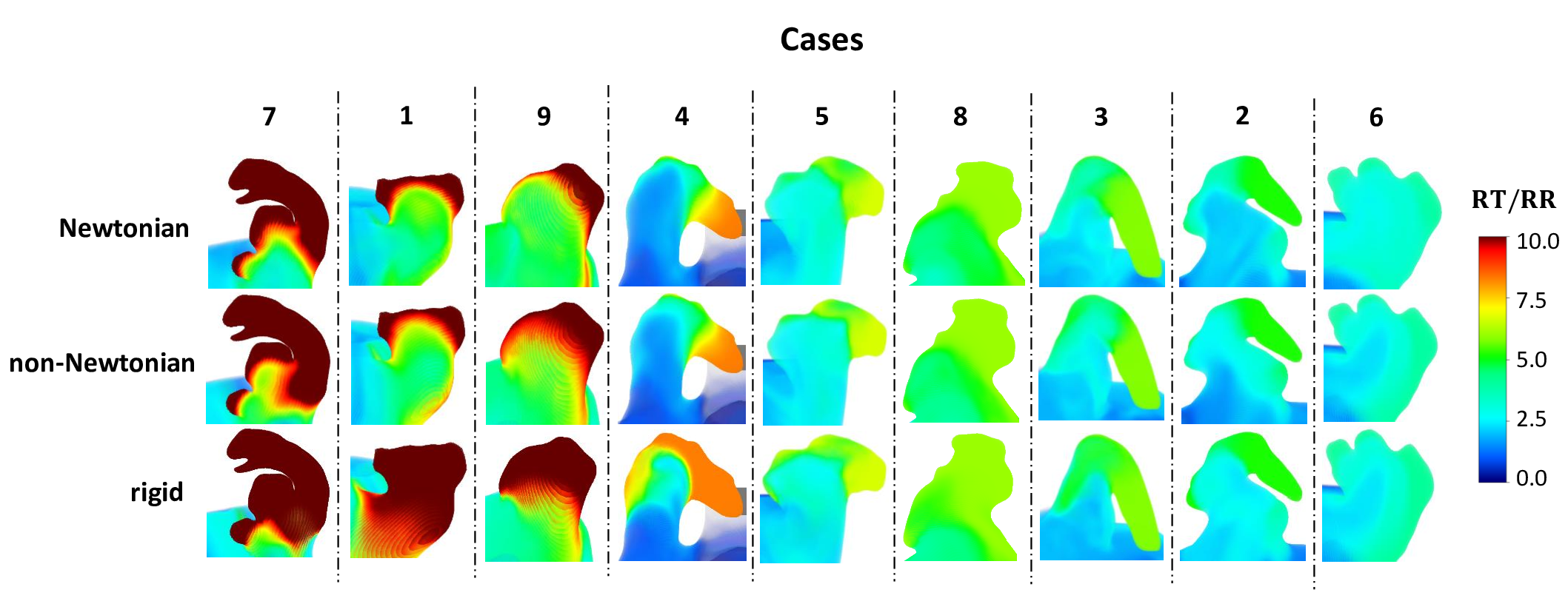}}
\caption{$\text{RT}/\text{RR}$ contours for the different models considered (rows). The cases are sorted in ascending HR.}
\label{fig:LAA_RT_by_time}
\end{figure*}

Another goal of the present \rebuttal{study} was to analyze the impacts of considering a Newtonian model and rigid walls separately. Figure \ref{fig:LAA_RT_by_time} shows \rebuttal{a} comparison of the $\text{RT}/\text{RR}$ contours for a uniform flow split (50-50\%), considering three cases: Newtonian wall motion model, non-Newtonian wall motion model, and non-Newtonian rigid wall model. The patients \rebuttal{are} displayed in different subcolumns in ascending order \rebuttal{of HR, and the} RT values \rebuttal{are} normalized with the patient-specific cardiac cycle (RR) duration.

As \rebuttal{shown} in Figure \ref{fig:LAA_RT_by_type}, the highest $\text{RT}/\text{RR}$ region is concentrated at the tip of \rebuttal{the} LAA. Nevertheless, the \rebuttal{extent} of this region depends on the model considered. \rebuttal{When} non-Newtonian models \rebuttal{were} compared, the LAA $\text{RT}/\text{RR}$ values increased in the rigid case. Interestingly, while the maximum extension of the $\text{RT}/\text{RR}$ region \rebuttal{was} markedly increased for the rigid wall in patients with \rebuttal{a} low HR, this increase \rebuttal{was} hardly visible in patients with \rebuttal{a} high HR. Figure \ref{fig:LAA_RT_by_time} \rebuttal{show the} $\text{RT}/\text{RR}$ for a Newtonian fluid and wall motion model. Unlike \rebuttal{the} non-Newtonian cases, \rebuttal{the} LAA $\text{RT}/\text{RR}$ values and the highest $\text{RT}/\text{RR}$ region \rebuttal{extensions} are lower.

\subsection{Stagnant blood volume evolution} 
\label{stag_vol_evolution}

\begin{figure*}[t!]
    \centering
    \begin{subfigure}[t]{0.46\textwidth}
        \centering
        \includegraphics[width=\linewidth]{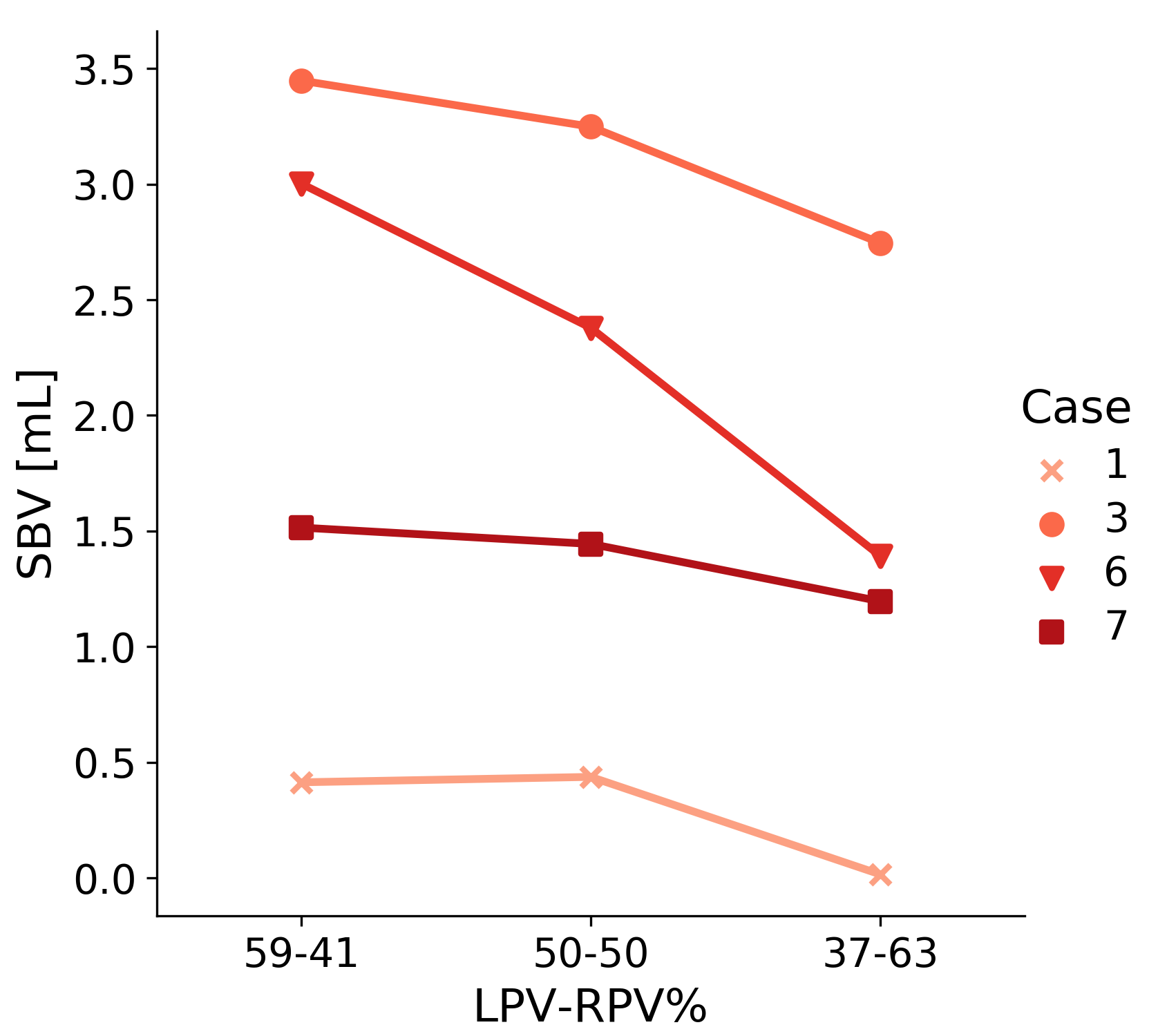}
        \caption{}
        \label{subfig:flowsplit_stagnant_volume_A}
    \end{subfigure}
    \begin{subfigure}[t]{0.46\textwidth}
        \centering
        \includegraphics[width=\linewidth]{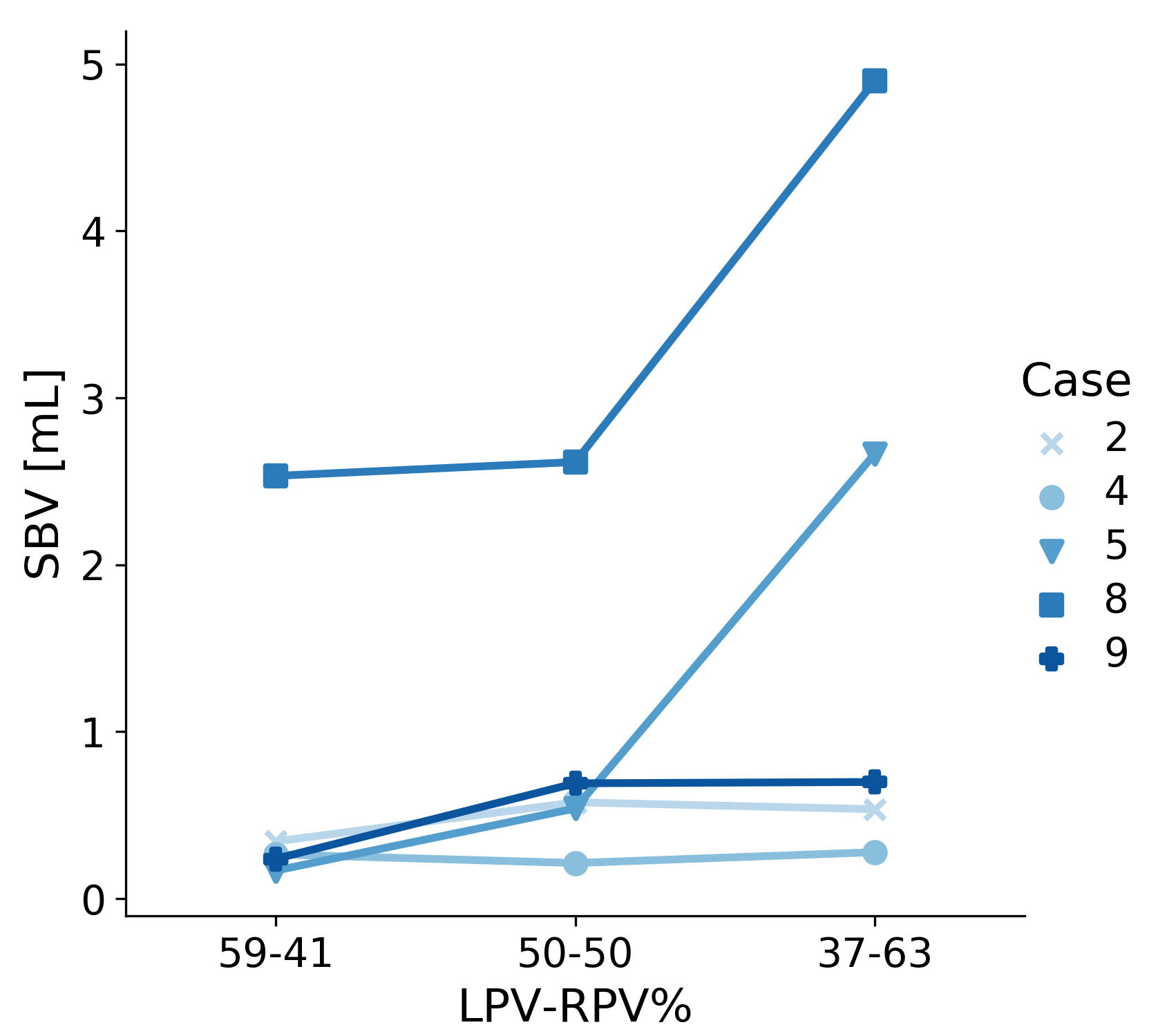}
        \caption{}
          \label{subfig:flowsplit_stagnant_volume_B}
    \end{subfigure}
    \begin{subfigure}[t]{0.46\textwidth}
        \centering
        \includegraphics[width=\linewidth]{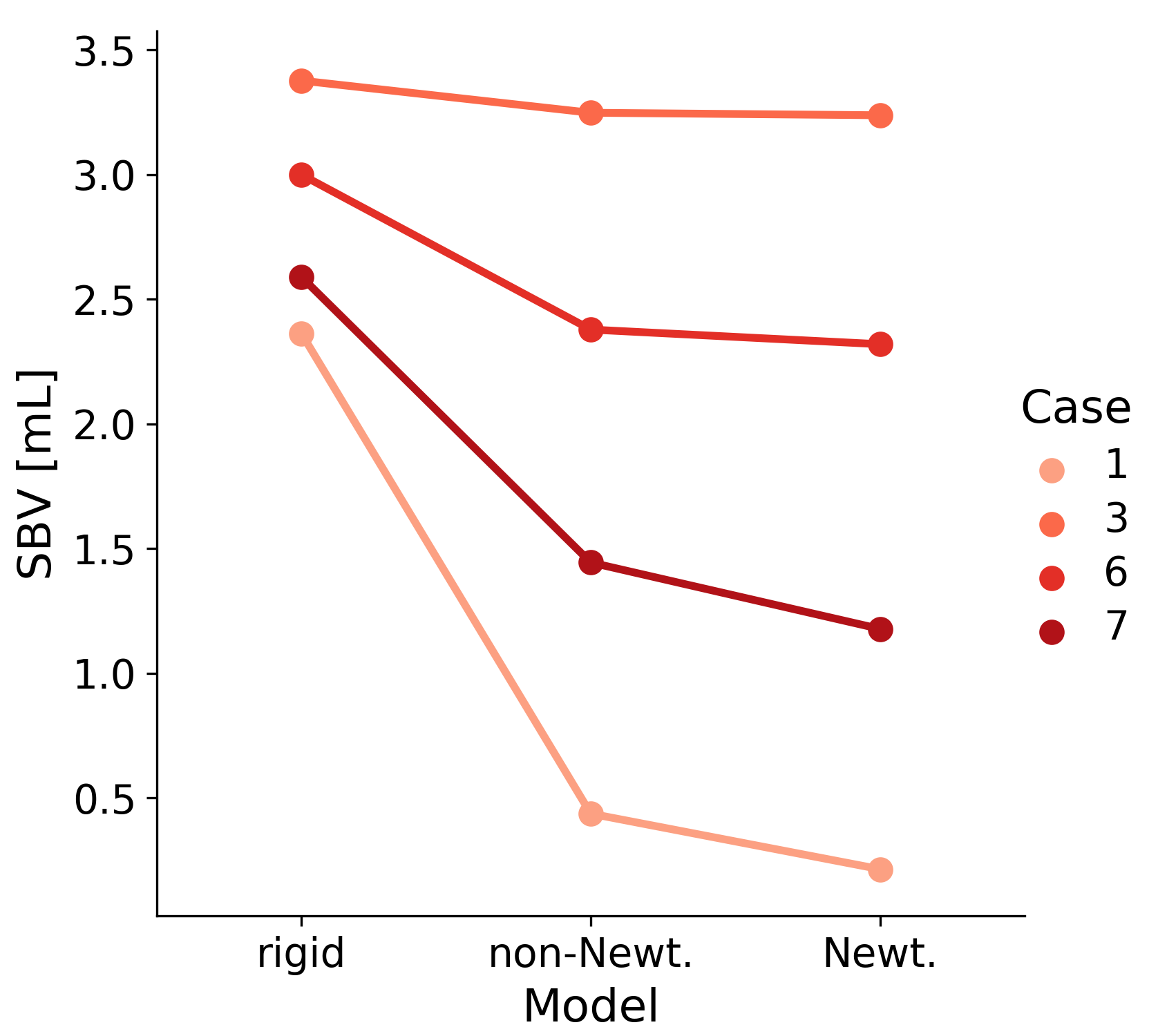}
        \caption{}
        \label{subfig:model_stagnant_volume_A}
    \end{subfigure}
    \begin{subfigure}[t]{0.46\textwidth}
        \centering
        \includegraphics[width=\linewidth]{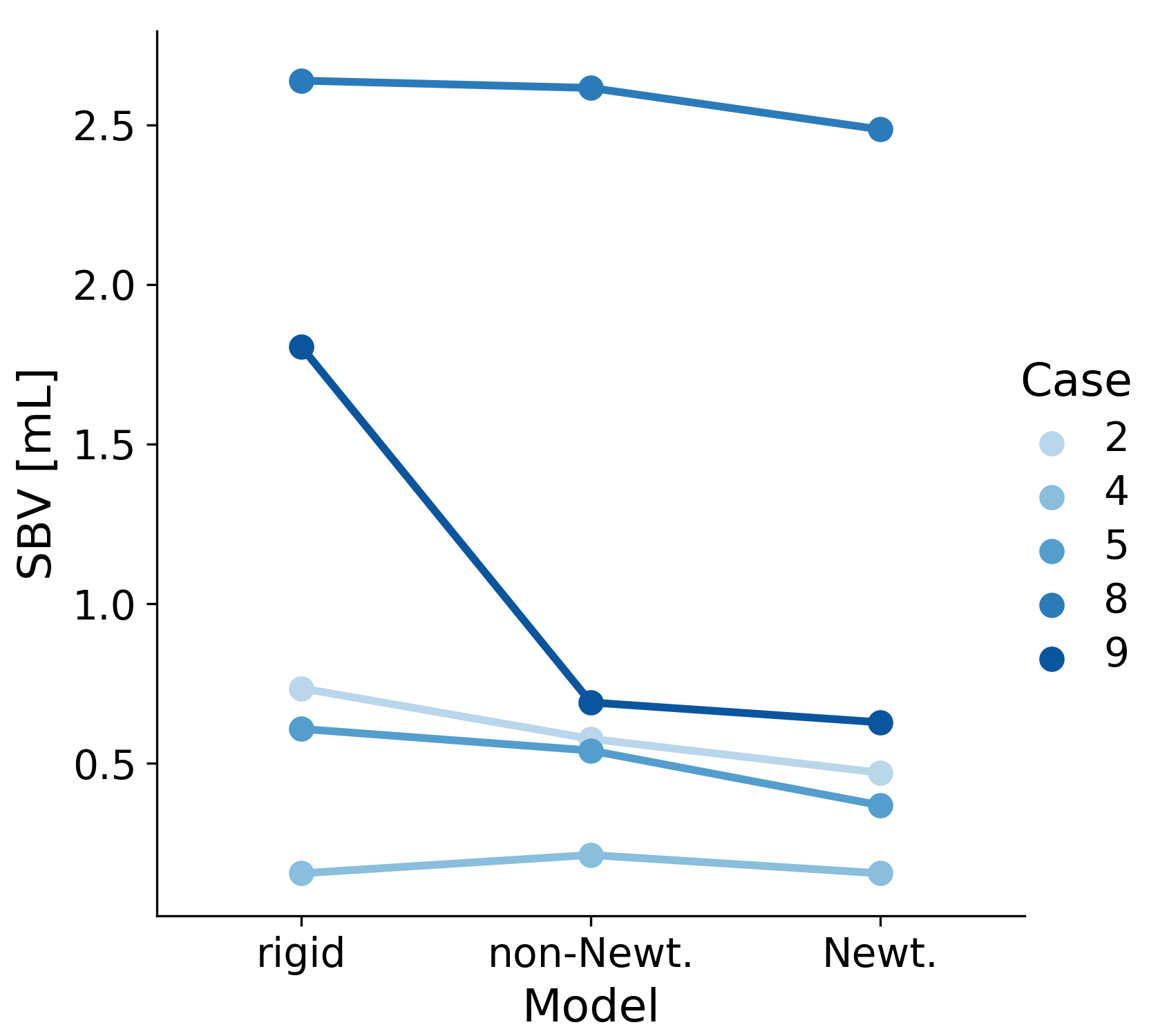}
        \caption{}
          \label{subfig:model_stagnant_volume_B}
    \end{subfigure}    
\caption{Comparison of the SBV for the different simulated cases. The first row shows the comparison depending on the flow split ratio: (a) Type A and (b) Type B, while the second row shows the comparison depending on the model: (c) Type A and (d) Type B. Types A and B are displayed in red and blue colors, respectively.}
\label{fig:stagnant_volume_evolution}
\end{figure*}

Following the procedure described in \rebuttal{Section} \ref{subsec:indices}, SBV was calculated for 45 simulations\rebuttal{, as} shown in Figure \ref{fig:stagnant_volume_evolution}, \rebuttal{and} Tables \ref{tab:stagnant_evolution_flow} and \ref{tab:stagnant_evolution_model}. The first row  in Figure \ref{fig:stagnant_volume_evolution} shows the differences in SBV between different flow splits (59-41, 50-50, 37-63\% LPV-RPV), while the second row illustrates the variations for the different models under consideration: non-Newtonian rigid wall model, non-Newtonian wall-motion model, and Newtonian wall-motion model. The first column shows the SBV for atrial type A \rebuttal{cases, which are} displayed in red. The second column depicts in blue the cases of type B. Tables \ref{tab:stagnant_evolution_flow} and \ref{tab:stagnant_evolution_model} show the numerical value of the SBV presented in Figure \ref{fig:stagnant_volume_evolution}, together with its relative variation ($\Delta$SBV). The models \rebuttal{used} as \rebuttal{references} were the 50-50\%LPV-RPV for the flow split ratio comparison and the \rebuttal{non}-Newtonian model for the \rebuttal{model} comparison. 

\begin{table}[htbp]
\centering
\caption{SBV variation for the different flow split ratios. The cases are classified as type A and type B, whose values are depicted in Figures \ref{subfig:flowsplit_stagnant_volume_A} and \ref{subfig:flowsplit_stagnant_volume_B}\rebuttal{, respectively}.} 
\begin{adjustbox}{max width=13cm}
\color{black}
\begin{tabular}{ccccc}
\textbf{Type}  & \textbf{Case}  & \textbf{\%LPV-RPV} & \begin{tabular}[c]{@{}c@{}}\textbf{SBV [mL]}\\ \end{tabular} & \begin{tabular}[c]{@{}c@{}}\textbf{$\Delta$SBV [\%]}\\ \end{tabular}  \\ 
\hline \multirow{12}{*}{\textbf{A}} & \multirow{3}{*}{1} & 59-41  & 0.41  & -6.8 \\   &       & 50-50  & 0.44   & 0.0   \\   &  & 37-63  & 0.02    & -95.5  \\ \cline{2-5} 
& \multirow{3}{*}{3} & 59-41   & 3.45  & 6.2  \\  &   & 50-50   & 3.25  & 0.0    \\  &   & 37-63   & 2.75  & -15.4 \\ \cline{2-5} 
& \multirow{3}{*}{6} & 59-41  & 3.00   & 26.1 \\ &  & 50-50  & 2.38   & 0.0   \\   &  & 37-63  & 1.39  & -41.6  \\ \cline{2-5}
& \multirow{3}{*}{7} & 59-41  & 1.51  & 4.9 \\
& 7 & 50-50  & 1.44 & 0.0  \\ 
&  & 37-63  & 1.20  & -16.7 \\ \hline
\multirow{15}{*}{\textbf{B}} & \multirow{3}{*}{2} & 59-41   & 0.34  & -41.4  \\ &       & 50-50  & 0.58  & 0.0   \\  & & 37-63  & 0.54  & -6.9  \\ \cline{2-5} & \multirow{3}{*}{4} & 59-41  & 0.27 & 28.6 \\ &   & 50-50  & 0.21   & 0.0  \\ &  & 37-63   & 0.28   & 33.3 \\ \cline{2-5} & \multirow{3}{*}{5} & 59-41  & 0.17  & -68.5 \\ &  & 50-50  & 0.54   & 0.0   \\ &  & 37-63  & 2.67  & 394.4  \\ \cline{2-5} & \multirow{3}{*}{8} & 59-41  & 2.53  & -3.4   \\  &   & 50-50  & 2.62   & 0.0    \\ &  & 37-63  & 4.90  & 87.0 \\ \cline{2-5}
 & \multirow{3}{*}{9} & 59-41  & 0.24   & -65.2 \\  &  & 50-50   & 0.69   & 0.0  \\ &  & 37-63  & 0.70  & 1.4 \\
\hline
\end{tabular}
\end{adjustbox}
\label{tab:stagnant_evolution_flow}
\end{table}

\begin{table}[htbp]
\centering
\caption{SBV variation for the different models. The cases are classified as type A and type B, whose values are depicted in Figures \ref{subfig:model_stagnant_volume_A} and \ref{subfig:model_stagnant_volume_B}\rebuttal{, respectively}.} 
\begin{adjustbox}{max width=13cm}
\color{black}
\begin{tabular}{ccccc}
\textbf{Type}                     & \textbf{Case}      & \textbf{Model} & \begin{tabular}[c]{@{}c@{}}\textbf{SBV [mL]}\\ \end{tabular} & \begin{tabular}[c]{@{}c@{}}\textbf{$\Delta$SBV [\%]}\\ \end{tabular}  \\ 
\hline
\multirow{12}{*}{\textbf{A}} & \multirow{3}{*}{1} & rigid   & 2.36  & 436.4          \\   &   & non-Newt.  & 0.44   & 0.0  \\  &  & Newt.  & 0.21  & -52.3 \\ 
\cline{2-5} & \multirow{3}{*}{3} & rigid  & 3.38  & 4.0   \\ &   & non-Newt.  & 3.25    & 0.0   \\ &   & Newt.  & 3.24  & -0.3   \\ \cline{2-5}  & \multirow{3}{*}{6} & rigid    & 3.00  & 26.1  \\   &  & non-Newt.  & 2.38   & 0.0   \\ &  & Newt. & 2.32   & -2.5  \\ \cline{2-5}   & \multirow{3}{*}{7} & rigid  & 2.59   & 79.9    \\    &                    & non-Newt.    & 1.44   & 0.0   \\  &   & Newt.  & 1.18    & -18.1   \\ 
\hline
\multirow{15}{*}{\textbf{B}} & \multirow{3}{*}{2} & rigid  & 0.74   & 27.6     \\
   &    & non-Newt.   & 0.58  & 0.0   \\  &  & Newt.    & 0.47     & -18.4     \\ 
\cline{2-5}  & \multirow{3}{*}{4} & rigid   & 0.16   & -27.6  \\  &  & non-Newt.          & 0.21   & 0.0  \\   &   & Newt.   & 0.15  & -28.6  \\ \cline{2-5} & \multirow{3}{*}{5} & rigid    & 0.61  & 13.0    \\  &     & non-Newt.   & 0.54     & 0.0   \\   &         & Newt.             & 0.39      & -27.8 \\ 
\cline{2-5}  & \multirow{3}{*}{8} & rigid     & 2.64     & 0.8  \\   &                    & non-Newt.          & 2.62     & 0.0    \\  &     & Newt.  & 2.49    & -5.0    \\ 
\cline{2-5} & \multirow{3}{*}{9} & rigid      & 1.81    & 162.3     \\  &                    & non-Newt.    & 0.69   & 0.0    \\  &   & Newt.  & 0.63    & -8.7   \\
\hline
\end{tabular}
\end{adjustbox}
\label{tab:stagnant_evolution_model}
\end{table} 

For type A \rebuttal{atria} (Figure \ref{subfig:flowsplit_stagnant_volume_A}), for which the main circulatory flow is located just after the LPV, \rebuttal{the} SBV tends to decrease as the flow split ratio in the RPV increases. It can also be seen that the magnitude of this variation depends on the individual (the SBV ranges between 0.4 and 3.4 mL for a \rebuttal{50-50\%}LPV-RPV). 

Regarding type B \rebuttal{atria} (Figure \ref{subfig:flowsplit_stagnant_volume_B}), the trend is the opposite: the SBV tends to increase as the flow split ratio in the RPV increases. This \rebuttal{occurred} smoothly, except for Cases 5 and 8, for which the increase \rebuttal{was} pronounced (Table \ref{tab:stagnant_evolution_flow}). 

The SBV \rebuttal{tended} to increase for the rigid wall model compared \rebuttal{with} the wall motion model. This effect \rebuttal{seemed} to be more pronounced for type A \rebuttal{atria} (Table \ref{tab:stagnant_evolution_model}). However, the differences in Cases 3, 5, and 8 \rebuttal{were negligible} when the wall motion model \rebuttal{was considered}. This general trend \rebuttal{was} reversed \rebuttal{in} Case 4, where the SBV \rebuttal{decreased} when the rigid wall model is considered. 

Regarding the non-Newtonian model cases, the SBV \rebuttal{tends} to be slightly higher than \rebuttal{that of} the Newtonian \rebuttal{cases}, which follows the same trend as \rebuttal{the} RT in Figure \ref{fig:LAA_RT_by_time}.

\subsection{Hemodynamic maps}
\label{subsec:hemodynamic_maps}

\begin{figure*}[t]
\centerline{\includegraphics[width=0.92\linewidth]{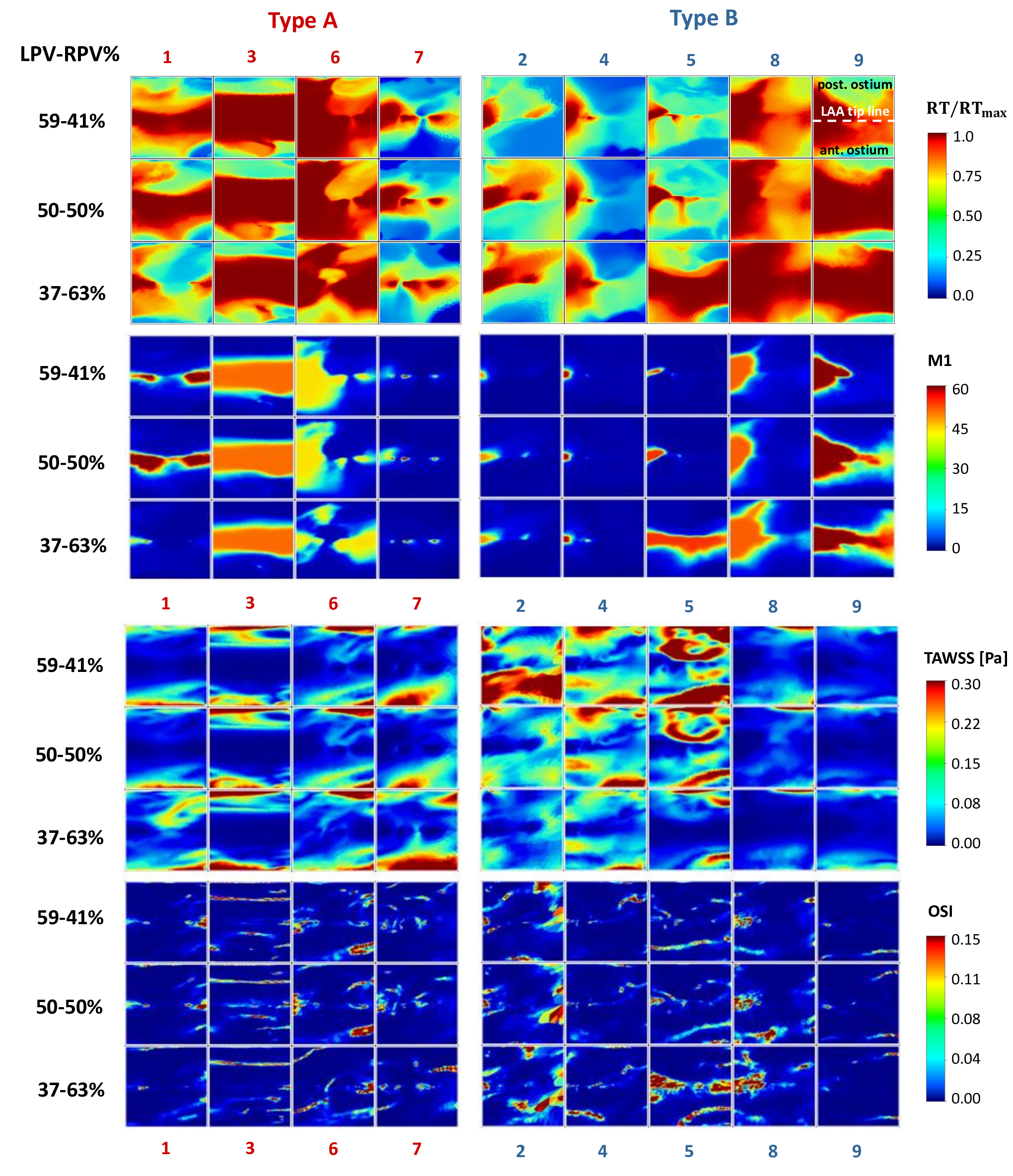}}
\caption{The hemodynamic variables resulting from the simulations were projected to the unit square by employing the ULAAC coordinates, facilitating 2D visualization and data comparison. Columns represent the nine different patients, while rows show the contours variation \rebuttal{because of} changes in the flow split ratio.}
\label{fig:ulaac_flowsplit}
\end{figure*}

The hemodynamic maps were calculated as described in \S  \ref{subsec:ulaac}. \rebuttal{The} ULAAC system was used to facilitate data visualization and comparison of results. Figure \ref{fig:ulaac_flowsplit} shows all hemodynamic indices introduced in \S \ref{subsec:indices}: $\text{RT}/\text{RT}_{max}$, M1, TAWSS and OSI, where RT was normalized dividing by $\text{RT}_{max}$ to facilitate visualization. The maps are organized in two main columns according to type A and type B, respectively. The subcolumns represent the different patients sorted as in Figure \ref{fig:LAA_RT_by_type}, while the rows illustrate the different flow split cases. Complementarily to the projections shown in Figure \ref{fig:ulaac_flowsplit}, Table \ref{tab:ulaac_flowsplit} presents both the percentage of LAA area (Area [\%]) exceeding a certain threshold (0.15 Pa for the TAWSS contours and 0.08 for OSI contours), and the difference ($\Delta$Area [\%]) with respect to the 50-50\% LPV-RPV area value.

The $\text{RT}/\text{RT}_{max}$ and M1 indices confirm the same trend as shown in \S \ref{subsec:LAA_RT} and \S \ref{stag_vol_evolution}. $\text{RT}/\text{RT}_{max}$ and M1 decrease (increase) as the RPV flow split increases for types A (B). In contrast, the opposite behavior can be seen in the TAWSS maps. Cases 3 and 9 notably depict this tendency (Table \ref{tab:ulaac_flowsplit}). Both cases show a wide region with high RT values concentrated near the LAA tip (Figure \ref{subfig:method_ulaac}). As the RPV flow split ratio increases, the TAWSS values in Case 3 increase, while in Case 9 decrease. Concerning OSI maps, the values are quite low, indicating reduced flow oscillations in all cases. However, as a general trend, OSI values decrease slightly (increase) as the RPV flow split increases for type A (B). This trend is also reflected in the variation of the percentage of LAA area shown in Table \ref{tab:ulaac_flowsplit}.

\begin{figure*}[t]
\centerline{\includegraphics[width=0.92\linewidth]{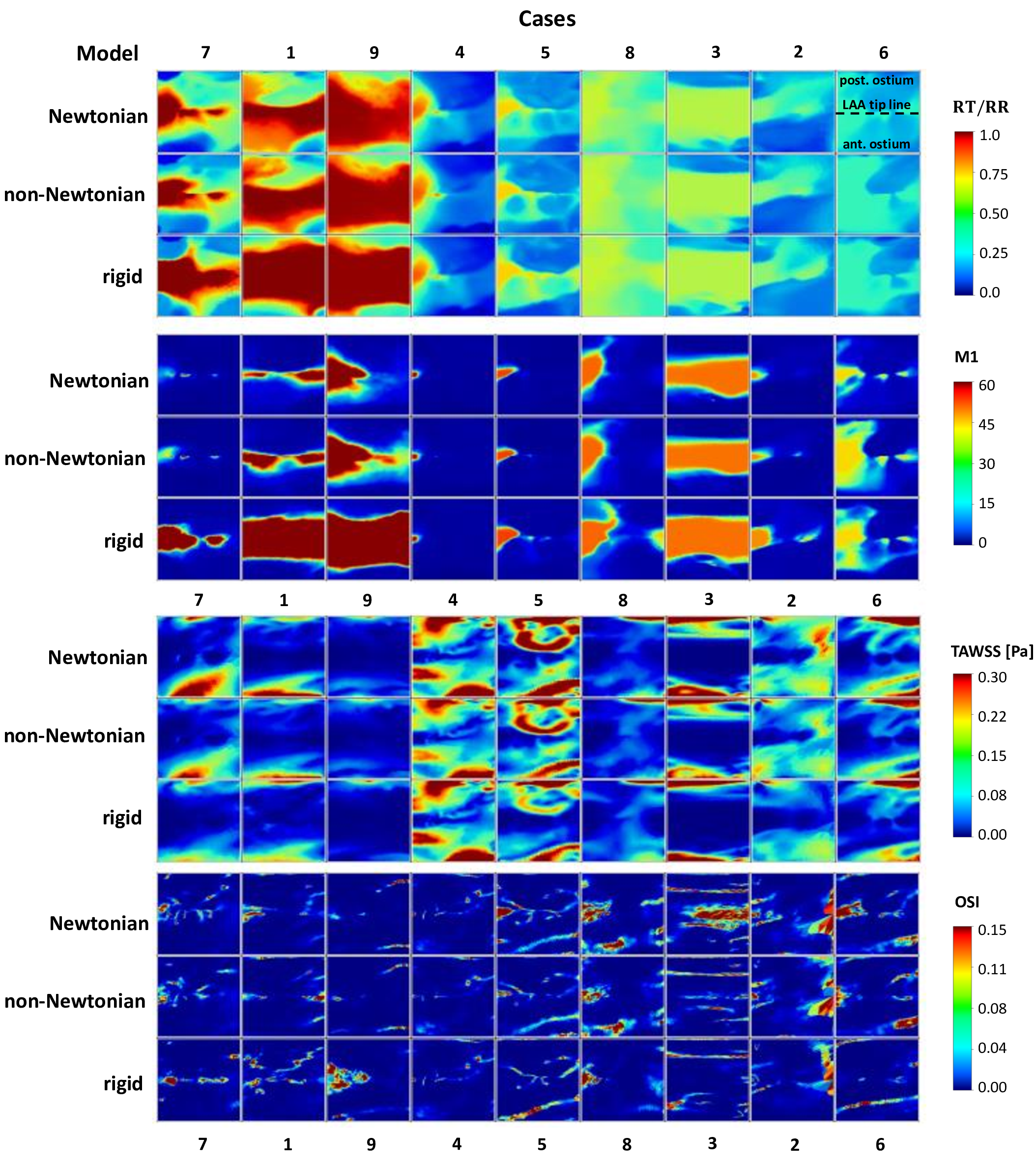}}
\caption{The hemodynamic variables resulting from the simulations were projected to the unit square by employing the ULAAC coordinates, facilitating 2D visualization and data comparison. The columns represent the nine different patients, while the rows give an appreciation of the variation of the contours \rebuttal{because of} the different models under consideration.}
\label{fig:ulaac_model}
\end{figure*}

Figure \ref{fig:ulaac_model} illustrates the hemodynamic maps for the three models under consideration: Newtonian wall-motion model, non-Newtonian wall-motion model, and non-Newtonian rigid wall model. The subcolumns show the patients in ascending HR order, while the rows illustrate the different models under consideration. As \rebuttal{shown} in Figure \ref{fig:LAA_RT_by_time}, RT values were normalized by dividing \rebuttal{them} by the patient-specific cardiac cycle duration (RR). Similar to Table \ref{tab:ulaac_flowsplit}, Table \ref{tab:ulaac_model} presents analogous data extracted from Figure \ref{fig:ulaac_model}. As in Table \ref{tab:ulaac_flowsplit}, Area [\%] express the percentage of LAA area over a certain threshold (0.15 Pa for the TAWSS contours and 0.08 for OSI contours). However, in this table, $\Delta$Area [\%] illustrates the difference with respect to the area calculated in the non-Newtonian model. 

The lowest HR patients (Cases 1, 7, and 9) show the greatest differences in all hemodynamic indices \ref{tab:ulaac_model}, while these differences \rebuttal{were} attenuated in \rebuttal{patients with} the highest HR. \rebuttal{The} Newtonian model slightly \rebuttal{improved} the renovation of blood within the LAA compared to non-Newtonian \rebuttal{models}. This reflects the same behavior \rebuttal{that described in Section} \ref{subsec:LAA_RT}.

\begin{table}[htbp]
\centering
\caption{Variation of the percent LAA area exceeding 
a certain threshold (0.15 Pa for TAWSS contours and 0.08 for OSI contours) for the different flow split ratios. The cases are classified as type A and type B, whose ULAAC \rebuttal{projections are} depicted in Figure \ref{fig:ulaac_flowsplit}.}  
\begin{adjustbox}{max width=11cm}
\color{black}
\begin{tabular}{ccccccc}
                    &     &   & \multicolumn{2}{c}{\textbf{TAWSS}}  & \multicolumn{2}{c}{\textbf{OSI}}   \\ \cline{4-7}
\textbf{Type}  &  \textbf{Case}    &  \textbf{\%LPV-RPV} & \begin{tabular}[c]{@{}c@{}}\textbf{Area}\\~{\textbf{[}\textbf{\%]}}\end{tabular} & \begin{tabular}[c]{@{}c@{}}\textbf{$\Delta$Area}\\~{\textbf{[}\textbf{\%]}}\end{tabular} & \begin{tabular}[c]{@{}c@{}}\textbf{Area}\\~\textbf{{[}\%]}\end{tabular} & \begin{tabular}[c]{@{}c@{}}\textbf{$\Delta$Area}\\~{\textbf{[}}\textbf{\%]}\end{tabular}  \\ 
\hline
\multirow{12}{*}{\textbf{A}} & \multirow{3}{*}{1} & 59-41     & 9.2     & 2.9      & 1.5         & -0.5     \\  &     & 50-50     & 6.3      & 0.0                              & 2.0          & 0.0     \\   &       & 37-63     & 16.6    & 10.3     & 0.8      & -1.2    \\ \cline{2-7}  & \multirow{3}{*}{3} & 59-41     & 20.3    & 0.2     & 4.4    & 0.7     \\  &     & 50-50     & 20.1    & 0.0    & 3.7    & 0.0        \\  &     & 37-63     & 17.0    & -3.1     & 2.9   & -0.8    \\ 
\cline{2-7}  & \multirow{3}{*}{6} & 59-41     & 9.5   & -2.1   & 5.0    & -0.7     \\ &     & 50-50     & 11.6  & 0.0   & 5.7     & 0.0   \\   &                    & 37-63     & 13.5   & 1.9    & 3.8    & -1.9   \\ 
\cline{2-7} & \multirow{3}{*}{7} & 59-41     & 17.8    & 6.4   & 1.3     & -0.3      \\   &    & 50-50     & 11.4  & 0.0    & 1.6   & 0.0    \\   &                    & 37-63     & 30.4      & 19.0   & 2.6     & 1.0     \\ 
\hline
\multirow{15}{*}{\textbf{B}} & \multirow{3}{*}{2} & 59-41     & 54.2    & 40.5     & 7.7      & -0.9      \\   &      & 50-50     & 13.7   & 0.0    & 8.6    & 0.0    \\ &                    & 37-63     & 2.1     & -11.6   & 10.4      & 1.8    \\ 
\cline{2-7}  & \multirow{3}{*}{4} & 59-41     & 40.0   & 0.4      & 1.4    & 0.7    \\  &      & 50-50     & 34.6   & 0.0     & 0.7    & 0.0     \\ &                    & 37-63     & 14.0       & -20.6     & 3.2      & 2.5    \\ 
\cline{2-7}  & \multirow{3}{*}{5} & 59-41     & 56.3   & 20.6     & 2.7    & 0.3  \\ &    & 50-50     & 35.7    & 0.0      & 2.4     & 0.0    \\   &                    & 37-63     & 10.8     & -24.9   & 13.9      & 11.5    \\ 
\cline{2-7}  & \multirow{3}{*}{8} & 59-41     & 6.1       & 3.7    & 2.7   & -3.4   \\ &     & 50-50     & 2.4  & 0.0    & 6.1    & 0.0    \\  &                    & 37-63     & 1.8    & -0.6     & 6.8      & 0.7       \\ 
\cline{2-7}  & \multirow{3}{*}{9} & 59-41     & 1.0      & 0.5     & 0.9     & -0.3    \\  &  & 50-50     & 0.5     & 0.0    & 1.2      & 0.0   \\   &                    & 37-63     & 1.5    & 1.0    & 0.8    & -0.4    \\
\hline
\end{tabular}
\end{adjustbox}
\label{tab:ulaac_flowsplit}
\end{table} 

\begin{table}[htbp]
\centering
\caption{Variation of the percent LAA area exceeding 
a certain threshold (0.15 Pa for TAWSS contours and 0.08 for OSI contours) for the different models selected. The cases are classified as type A and type B, whose ULAAC \rebuttal{projections are} depicted in Figure \ref{fig:ulaac_model}.}
\begin{adjustbox}{max width=11cm}
\color{black}
\begin{tabular}{ccccccc}
                    &     &     & \multicolumn{2}{c}{\textbf{TAWSS}}  & \multicolumn{2}{c}{\textbf{OSI}}    \\ \cline{4-7} \textbf{Type}                & \textbf{Case}               & \textbf{Model}  & \begin{tabular}[c]{@{}c@{}}\textbf{Area}\\~{\textbf{[}}\textbf{\%]}\end{tabular} & \begin{tabular}[c]{@{}c@{}}\textbf{$\Delta$Area}\\~{\textbf{[}}\textbf{\%]}\end{tabular} & \begin{tabular}[c]{@{}c@{}}\textbf{Area}\\~{\textbf{[}}\textbf{\%]}\end{tabular} & \begin{tabular}[c]{@{}c@{}}\textbf{$\Delta$Area}\\~{\textbf{[}}\textbf{\%]}\end{tabular}  \\ 
\hline
\multirow{12}{*}{\textbf{A}} & \multirow{3}{*}{1} & rigid      & 9.9     & 3.6     & 3.0        & 0.9       \\  &     & non-Newt. & 6.3     & 0.0     & 2.1    & 0.0             \\  &    & Newt.    & 13.8     & 7.5    & 1.7      & -0.4        \\ 
\cline{2-7}  & \multirow{3}{*}{3} & rigid      & 14.4    & -5.7    & 4.4      & 0.7      \\  &     & non-Newt. & 20.1   & 0.0      & 3.7      & 0.0   \\  &                    & Newt.    & 24.5    & 4.4   & 14.4     & 10.7       \\ 
\cline{2-7}  & \multirow{3}{*}{6} & rigid      & 15.2     & 3.6     & 1.7    & -4.0    \\ &   & non-Newt. & 11.6    & 0.0      & 5.7      & 0.0    \\
                    &   & Newt.    & 16.5     & 4.9     & 5.8   & 0.1      \\ 
\cline{2-7}   & \multirow{3}{*}{7} & rigid      & 4.0      & -7.4      & 2.6    & 1.0     \\   &     & non-Newt. & 11.4    & 0.0     & 1.6     & 0.0      \\
                    &       & Newt.    & 20.9      & 9.5   & 2.0     & 0.4    \\ 
\hline
\multirow{15}{*}{\textbf{B}} & \multirow{3}{*}{2} & rigid      & 8.1    & -5.6      & 5.0     & -3.6    \\   &      & non-Newt. & 13.7    & 0.0    & 8.6     & 0.0    \\  &                    & Newt.    & 25.5     & 11.8      & 7.8     & -0.8   \\ 
\cline{2-7}  & \multirow{3}{*}{4} & rigid      & 40.6    & 6.0   & 0.8    & 0.1     \\   &    & non-Newt. & 34.6  & 0.0    & 0.7    & 0.0     \\   &                    & Newt.    & 40.6      & 6.0     & 0.8      & 0.1    \\ 
\cline{2-7}  & \multirow{3}{*}{5} & rigid      & 24.7    & -11.0    & 4.4     & 2.0      \\   &      & non-Newt. & 35.7     & 0.0     & 2.4        & 0.0      \\
                    &      & Newt.    & 41.0      & 5.3       & 3.9     & 1.5     \\ \cline{2-7}  & \multirow{3}{*}{8} & rigid      & 3.6      & 1.1   & 2.3     & -3.8   \\  &     & non-Newt. & 2.5        & 0.0    & 6.1      & 0.0    \\
                    &     & Newt.    & 3.5    & 1.0       & 7.7    & 1.6           \\ 
\cline{2-7}
                    & \multirow{3}{*}{9} & rigid      & 0.8     & 0.3    & 5.8      & 4.6   \\  &    & non-Newt. & 0.5   & 0.0     & 1.2   & 0.0    \\  &     & Newt.    & 0.4    & -0.1      & 1.1  & -0.1      \\
\hline
\end{tabular}
\end{adjustbox}
\label{tab:ulaac_model}
\end{table}

\section{Discussion}
\label{sec:discussion}

In recent years, \rebuttal{the} CFD analysis of LA blood flow \rebuttal{in} patient-specific models has been recognized as an essential tool \rebuttal{for deepening} our understanding of the hemodynamic substrate of stroke in patients with AF \cite{corti2022impact, duenas2021comprehensive, Garcia-Isla2018, garcia2021demonstration}. Using endocardial geometries from medical imaging, prescribed combinations of flow boundary conditions \rebuttal{at} the inlets/outlets, and wall motion reconstruction, CFD simulations can provide an accurate 3D time-resolved atrial flow representation. This time-resolved flow field allows \rebuttal{the} calculation of surrogate hemodynamic metrics related to stasis \cite{duenas2022morphing, gonzalo2022non,pons2022joint}, such as blood age or wall shear-derived indices (\rebuttal{e.g., TAWSS and} OSI). These metrics are especially relevant within the LAA \rebuttal{because} it is the most prothrombotic of intracardiac regions. 

Previous CFD studies \rebuttal{have} focused on the relationship between LAA morphology and stasis \cite{back2023, duenas2022morphing, Garcia-Isla2018, garcia2021demonstration, Masci2019}, proposing new metrics to quantify stasis \cite{corti2022impact, Achille2014, duenas2021comprehensive}, or contrasting different model assumptions for LA CFD simulations \cite{duenas2021comprehensive, duenas2021boundary,duran2023pulmonary,gonzalo2022non, Khalili2024,Mill2021}. However, the formation of a SBV within the LAA is governed by atrial flow patterns \cite{back2023}. \rebuttal{Atrial} flow patterns are sensitive to various parameters \rebuttal{such as}: LAA morphology and relative position, PV orientations and flow split ratio, transmitral flow profile, and LA wall motion\rebuttal{, etc}. Therefore, the scientific community faces a multifactorial problem that cannot be easily solved by studying the \rebuttal{effects} of a single parameter. For these reasons\rebuttal{,} and despite the efforts made in recent years, the relationship between LA flow patterns, LA/LAA morphology, and \rebuttal{LAA stasis} is not yet fully understood. Although these studies have helped better understand the phenomena of AF and select the boundary conditions and numerics that best reproduce atrial flow, the published analyses have not yet \rebuttal{been sufficient} to build a predictive understanding of their anatomical and functional determinants. In addition, some of these variables can vary significantly over time \rebuttal{in} the same patient. Atrial remodeling and fibrosis can exacerbate the hemodynamic substrate of thrombosis over months/years \cite{boyle2021fibrosis}, while other variables such as the transmitral flow profile and the motion of the LA wall vary greatly between rest/exercise and sinus/fibrillated conditions \cite{Elliott2023,FARESE2019,TAKAGI2012}. Moreover, the flow split ratio \rebuttal{significantly} depends on the left/right lying position of the patient \cite{duran2023pulmonary,WRAPU19}. 

To meet the challenges mentioned above, this manuscript \rebuttal{combined} some of the most recent methodological advances \rebuttal{elucidate} the relationship between atrial flow patterns and stasis within the LAA. All these methods \rebuttal{have been} validated and used in previous \rebuttal{studies. T}he LAA wall motion model, which we used to normalize fibrillated motion between different patients, was proposed by Zíngaro \textit{et al.} \cite{zingaro2021geometric} and utilized in \cite{corti2022impact,zingaro2021hemodynamics}. Our group recently validated the \rebuttal{ULAAC} projection method \cite{duenas2024reduced} to facilitate data visualization and comparison between patients, and the SBV calculation method \rebuttal{has been} shown \rebuttal{to be} as an efficient way to delineate the stagnant region in previous \rebuttal{studies} \cite{duenas2021comprehensive, duenas2022morphing}.
The results highlight how the flow split ratio has a considerable influence on the atrial flow patterns of some patients but not for others. This evidence suggests that the influence of the flow split ratio on atrial flow patterns is governed by LA/LAA morphology and thus can be a determinant of the patient-specific risk of LAA stasis. Finally, some of the assumptions commonly used in atrial numerical models \rebuttal{were} tested to study their influence on these determinant atrial flow patterns\rebuttal{, including the evaluation} of the Newtonian fluid assumption and comparison of different wall motion models. 

First, in an effort to overcome the number of patients in our previous studies, nine patients with AF were imaged and segmented. Although we acknowledge that the number of patients is still limited, the number is significantly higher than \rebuttal{the state-of-the-art} (e.g., García \textit{et al.} \cite{garcia2021demonstration} performed 12 simulations on six patient-specific geometries), and the patient cohort was diverse enough to study and classify the different atrial flow patterns. Furthermore, the cohort \rebuttal{contained} not only atrial geometries but also PV and MV Doppler velocities, \rebuttal{allowing} the imposition of patient-specific boundary conditions. Therefore, performing 45 simulations on nine patient-specific geometries is a more significant number of simulations than most previous CFD atrial studies \cite{Bosi2018,duenas2021comprehensive,gonzalo2022non, garcia2021demonstration,Koizumi2015, Otani2016}. 

Depending on the patient-specific spatial distribution of atrial fibrosis, wall movement can be locally affected during \rebuttal{AF} episodes, which may introduce bias \rebuttal{into} our analysis. To homogenize and isolate this effect between different patients, we implemented a kinematic model capable of reproducing the variation in atrial volume according to the specific transmitral flow profile of the patient  \cite{corti2022impact,zingaro2021hemodynamics,zingaro2021geometric}. This kinematic model allowed us to generate smooth wall motion according to patient-specific MV Doppler measurements.  

As it is well known, LA flow is characterized by a circulatory flow induced by the flow coming from the LPV, while the flow from the RPV forms a stream directed \rebuttal{toward} the MV \cite{Vedula2015}. As one of the objectives of this \rebuttal{study} was to clarify how atrial flow patterns affect \rebuttal{the} hemodynamics within the LAA, the results obtained for different patients were classified according to the relative position of \rebuttal{the} main circulatory flow within the LA. \rebuttal{Thus}, three different flow behaviors were detected\rebuttal{, namely,} type A, B1, and B2. We classified the atrial flow pattern as type A when the main circulatory flow forms \rebuttal{immediately} after the LA entrance and near the LPV, type B1 when it forms downstream near the ostium, and type B2 when it is pushed toward the LA roof and near the RPV. Representative cases of each type are shown in Figure \ref{fig:velocity_streamlines}, \rebuttal{in which} additional views \rebuttal{are} added to facilitate circulatory flow visualization.

Interestingly, we noticed that the three different types of atrium A, B1, and B2\rebuttal{,} had different LAA washing behaviors and, even more \rebuttal{importantly}, different responses to changes in the flow split ratio. \rebuttal{We found} that the relative position of the main atrial circulatory flow \rebuttal{was} affected by the flow split ratio, which, to the best of our knowledge, \rebuttal{has} not yet been explored. This could have direct implications on the patient's lifestyle and health, as the position regularly adopted while sleeping (\rebuttal{left/right} lying position) changes the flow split ratio, \rebuttal{thus}affecting the position of the main atrial circulatory flow and LAA stasis risk in the medium and \rebuttal{long term}.

According to previous work \cite{duran2023pulmonary, gonzalo2022non}, we verified that the proximity of the main atrial circulatory flow to the ostium can induce secondary swirling \rebuttal{flow} within the LAA\rebuttal{, facilitating} washing and \rebuttal{preventing} blood stagnation. However, as \rebuttal{previously} mentioned, this position can be significantly modified by \rebuttal{changing} the flow split ratio. The effect of changing the flow split ratio strongly \rebuttal{depended} on the initial position of the main circulatory flow. For example, for an atrium with the main circulatory flow located \rebuttal{immediately} after the LPV (Type A), the effect of changing the flow split ratio from \rebuttal{59-41\%} to \rebuttal{37-63\%} can displace the main circulatory flow downstream toward the ostium, increasing its influence in the region near the LAA (Figure \ref{subfig:streamlines_A}). This proximity \rebuttal{induces} more robust LAA secondary flows\rebuttal{,} which \rebuttal{improve} washing, as can be seen from the decrease in RT (Figures \ref{fig:LAA_RT_by_type} and \ref{fig:ulaac_flowsplit}) and SBV (Figure \ref{subfig:flowsplit_stagnant_volume_A}, Table \ref{tab:stagnant_evolution_flow}). We found the opposite trend for \rebuttal{the} atrium with the main circulatory flow located near the ostium (Type B1). For type B1, LAA washing is already governed by LPV flow (Figure \ref{subfig:streamlines_B1}), so changing the flow split ratio from \rebuttal{59-41\%} to \rebuttal{37-63\%} only decreases the flow from the LPVs and thus increases RT (Figures \ref{fig:LAA_RT_by_type} and \ref{fig:ulaac_flowsplit}) and SBV (Figure \ref{subfig:flowsplit_stagnant_volume_B}, Table \ref{tab:stagnant_evolution_flow}). As before, the flow from the RPVs runs close to the atrial wall\rebuttal{: therefore,} LAA washing depends entirely on the flow coming from the LPVs and the position of the main atrial circulatory flow. 

On the other hand, the flow behavior is \rebuttal{completely} different for Type B2 \rebuttal{atria}, for which the main atrial circulatory flow is pushed against the atrial roof near the RPV. This results in a more chaotic atrial flow (Figure \ref{subfig:streamlines_B2}), which, to the best of our knowledge, has not been described before. Here, the RPV flow has two possible configurations depending on the flow split ratio: it can either collide with \rebuttal{or surround} the main atrial vortex. If the RPV flow surrounds the main atrial circulatory flow, this results in an exceptional situation in which washing in \rebuttal{the} LAA is governed by the RPV flow, with the RT decreasing as the RPV flow increases (Figure \ref{fig:LAA_RT_by_type}). This atrial flow behavior \rebuttal{has not been} previously observed. 

The same LAA washing patterns found when analyzing the streamlines and RT \rebuttal{were} reflected by other hemodynamic variables such as M1, TAWSS, and OSI, projected onto the ULAAC system \rebuttal{as shown} in Figure \ref{fig:ulaac_flowsplit}. It can be seen that the regions on the LAA surface with the highest M1 values are also associated with the highest RT and lowest TAWSS values \cite{duenas2021comprehensive, duenas2024reduced}. These prothrombotic regions tend to \rebuttal{be concentrated} near the LAA tip. Higher TAWSS values \rebuttal{tended} to concentrate on the borders of this stagnant region, although \rebuttal{their} distribution \rebuttal{seemed} to depend greatly on patient-specific LAA morphology. The areas calculated in Table \ref{tab:ulaac_flowsplit} helped verify the different trends for the Type A and Type B \rebuttal{atria} while varying the flow split ratio. 

Regarding the analysis of the vortex structures, the $q_2$ criterion was calculated and displayed for $q_2 = 0.2$ (Figure \ref{fig:vor_vortex_evo}). The $q_2$ criterion is widely used to assess flow characteristics and visualize vortex structures \cite{Hunt1988,Jeong1995}, also for atrial flow \cite{chnafa2014image, Otani2016}. This analysis helped confirm the vortex positions for each type of atrium, as previously described by the atrial streamlines. Some representative time instants were selected for each patient to show the temporal evolution of atrial flow structures (Figure \ref{fig:vor_vortex_evo}). Three different vortical patterns were identified \rebuttal{based on} the position of the main atrial vortex (Types A, B1, and B2). Once the MV is open, the higher vorticities are located \rebuttal{immediately} after the LPV for type A \rebuttal{atria, whereas} for type B1\rebuttal{,} they are displaced downstream near the ostium. It can also be seen that between 80-100\%RR, the main atrial vortex is fully developed. \rebuttal{In contrast}, the flow \rebuttal{in} type B2 \rebuttal{atria} presents a more chaotic pattern \rebuttal{owing} to the collision between the RPV and LPV flows, which produces high\rebuttal{-}magnitude vorticities. These observations \rebuttal{confirmed the previously observed} flow patterns, allowing us to \rebuttal{better} understand the relationship between the position of the main atrial vortex and LAA washing.

Regarding the comparison between \rebuttal{the} rigid-wall and moving-wall models, the general trend \rebuttal{was} an increase in \rebuttal{the} SBV when there \rebuttal{was} no wall motion (Figures \ref{subfig:model_stagnant_volume_A} and \ref{subfig:model_stagnant_volume_B}, Table \ref{tab:stagnant_evolution_model}). This result is consistent with previous work suggesting that the rigid model, although it does not always reflect patient-specific atrial flow, could be a conservative approximation for patients with impaired atrial wall motion \cite{duenas2021estimation, duenas2021comprehensive, garcia2021demonstration}. Interestingly, we found that this behavior \rebuttal{was} influenced by patient-specific HR (Figures \ref{fig:LAA_RT_by_time} and \ref{fig:ulaac_model}). The rigid model seems to overestimate RT for patients with \rebuttal{a} low HR, \rebuttal{whereas} it provides better approximation to the wall motion model for patients with \rebuttal{a} high HR. A possible explanation for \rebuttal{the} low HR is that the motion of the LAA wall \rebuttal{strongly} influences LAA washing. \rebuttal{In contrast}, in patients with \rebuttal{a} higher HR, LAA washing \rebuttal{was} governed by LA flow patterns. To \rebuttal{the best of} our knowledge, this effect \rebuttal{has} not previously been observed, as numerical studies tend to study and compare the results in a patient cohort for the same HR \cite{corti2022impact,duran2023pulmonary,garcia2021demonstration,Masci2019, zingaro2021hemodynamics}. As reported in previous studies \cite{duenas2021comprehensive, garcia2021demonstration}, we confirmed that the differences between the \rebuttal{rigid and the motion wall} models \rebuttal{showed} high variability \rebuttal{among} patients. This difference is reduced in patients with impaired atrial motion, even showing a lower SBV for the rigid model in some cases (see \rebuttal{Patient} 4 in Figure \ref{subfig:model_stagnant_volume_B} and Table \ref{tab:stagnant_evolution_model}), highlighting the multifactorial nature of this problem.

This underestimation of LAA washing observed for the rigid model when studying RT is also reflected by other hemodynamic variables such as M1, TAWSS\rebuttal{,} and OSI, which are displayed in Figure \ref{fig:ulaac_model}. \rebuttal{Similar to the} simulations that compare different flow split ratios, high\rebuttal{-}RT regions on the LAA surface \rebuttal{were} associated with high M1 values and low TAWSS values. \rebuttal{The} OSI \rebuttal{tended} to concentrate on the borders of stagnant regions, although its distribution \rebuttal{was} very irregular and \rebuttal{seemed} to depend on the particularities of each patient-specific geometry. The areas calculated in Table \ref{tab:ulaac_model} \rebuttal{help} verify the different TAWSS and OSI trends for the different models under consideration. 

A constitutive \rebuttal{relationship} based on the Carreau model was used \rebuttal{to compare} the Newtonian and non-Newtonian models. This \rebuttal{relationship has been} widely used in previous CFD studies of cardiovascular \rebuttal{flow} \cite{Agujetas2018, Albors2023rheological, gonzalo2022non}. However, the influence of non-Newtonian atrial effects \rebuttal{has} barely \rebuttal{been} studied compared to other parts of the cardiovascular system \rebuttal{because} for cardiac cavities, the Newtonian flow assumption is widely used as non-Newtonian atrial effects are reduced \cite{Garcia-Isla2018,Mill2020}. However, some recent \rebuttal{studies} \cite{Gonzalo22,Sanatkhani2023,Zhang2023} suggested that Newtonian effects can affect hemodynamics within the LAA, shaping both filling and draining jets and secondary swirling motions that govern LAA washing when \rebuttal{the} LA wall motion is impaired. We \rebuttal{confirmed} these observations in our patient cohort, showing \rebuttal{that considering} non-Newtonian effects implies slightly higher values of LAA RT, especially near the LAA tip (Figures \ref{fig:LAA_RT_by_time} and \ref{fig:ulaac_model}). This increase in stasis \rebuttal{indicated} an increase in SBV (Figures \ref{subfig:model_stagnant_volume_A} and \ref{subfig:model_stagnant_volume_B}, Table \ref{tab:stagnant_evolution_model}), which \rebuttal{increased} the prothrombotic substrate within the LAA. The results suggest that the Newtonian model slightly underestimates \rebuttal{the} RT in the LAA\rebuttal{; however,} as the results do not differ \rebuttal{significantly}, it can still be a good approximation.

\subsection{Clinical applications}
\label{subsec:clinical}

Recent studies have \rebuttal{demonstrated} the value of using CFD techniques to \rebuttal{better} understand cardiovascular \rebuttal{diseases} \cite{Rigatelli2022} and assist cardiac diagnosis \cite{Morris2016}. In the case of atrial fibrillation (AF), these techniques contributed to a better understanding of atrial flow patterns \cite{corti2022impact,duenas2021comprehensive, Garcia-Isla2018, garcia2021demonstration} and help quantify stasis within the LAA of a particular patient \cite{duenas2021comprehensive,Gonzalo22,Mill2020}. This could pave the way for \rebuttal{the creation of} diagnostic tools that \rebuttal{can} help identify groups of patients \rebuttal{and} choose the best treatment. However, although these CFD studies have helped to \rebuttal{better} understand the mechanisms of AF and to calculate parameters related to blood stasis that would otherwise be inaccessible, the mechanistic link between atrial flow patterns and stasis within \rebuttal{the} LAA is not fully understood. This study \rebuttal{investigated} atrial flow patterns, shedding light on the relationship between the position of the main atrial vortex and its implications for LAA stasis. 

Moreover, \rebuttal{based on the} flow split ratio analysis, we suggest that the fact that a patient recurrently sleeps on \rebuttal{the} left/right side may affect the risk of thrombosis in the medium and long term. We also showed how the performance of patient-specific simulations can determine which side is more favorable for \rebuttal{sleeping} and whether this will significantly impact the risk of thrombosis \rebuttal{in} this particular patient.

\subsection{Limitations}
\label{subsec:limitations}

This manuscript features nine patient-specific geometries with particular PV and MV Doppler-measured velocities, which \rebuttal{indicate} a significant increase in cohort size \rebuttal{compared to} most studies in this field. However, a larger cohort \rebuttal{is} necessary to perform statistical tests and draw general conclusions.

\rebuttal{The patient specificity} of the presented model may be limited by \rebuttal{the use of} an atrial kinematic model, even if it is derived on a patient-specific basis. Other studies have obtained atrial movement by registering and interpolating several endocardial images during the cardiac cycle \cite{duenas2021comprehensive,garcia2021demonstration,Otani2016}. However, \rebuttal{because} physicians tend to limit radiation exposure to patients, the application of a kinematic model \cite{corti2022impact,zingaro2021hemodynamics,zingaro2021geometric} offers a useful alternative. In fact, using smooth motion allowed us to isolate the effect of local fibrotic patterns \rebuttal{because of} AF from the anatomical effects, facilitating the comparison between different endocardial models.

\rebuttal{Similar} to most previous CFD studies in this field \cite{bucelli2022mathematical,duenas2021comprehensive, garcia2021demonstration,Otani2016}, the LV and the MV were not considered in the atrial model. This assumption is reasonable \rebuttal{because the simulation} of the entire left heart \rebuttal{suggests} that LV and MV dynamics do not significantly affect the LA flow structures \cite{Vedula2015}.

Although biochemical modeling of
thrombogenesis could be incorporated into our blood analysis to model the coagulation cascade and subsequent thrombus formation, it would have required considering multiple biochemical equations \cite{RMN18, seo2016coupled}, considerably increasing the computational time. Furthermore, the \rebuttal{patient specificity} of these models is still limited \cite{garcia2021demonstration}.

The orientation of the PVs was not modified \rebuttal{and} patient-specific PV orientations \rebuttal{were maintained}. However, recent studies \rebuttal{have pointed out} their influence on LAA stasis \cite{duenas2022morphing, mill2023role}, suggesting that they influence the position where the main atrial vortex is formed. We intend to address this \rebuttal{issue} in our future work \rebuttal{because the high} computational \rebuttal{costs} involved. 

\section{Conclusion} 
\label{subsec:conclusion}

This study aimed to achieve a deeper understanding of the relationship between atrial flow patterns and stasis within the LAA, with a particular focus on the formation and position of the main vortex that governs atrial flow. To this end, three of the most recent methodological advances were combined: a kinematic atrial motion model to isolate wall motion from \rebuttal{geometric} effects \cite{corti2022impact,zingaro2021hemodynamics,zingaro2021geometric}, a projection of the wall results on the ULAAC system to improve visualization \cite{duenas2024reduced}, and \rebuttal{the} calculation of the SBV to quantify stasis within the LAA \cite{duenas2021comprehensive, duenas2022morphing}. Although atrial vortex structures have been previously observed \cite{Masci2019, Otani2016} and suggested to condition stasis within the LAA, our results highlight that both the patient-specific geometry of the atrium and the flow split ratio have an important influence on the position of \rebuttal{the} main atrial flow vortex. To the best of our knowledge, this has not been \rebuttal{previously} observed.

Furthermore, the effect of the flow split ratio on the position of the main atrial vortex \rebuttal{is} essential for some patients and negligible for others, showing high patient-specific variability. This variable could have important clinical implications\rebuttal{,} as the flow split ratio is severely affected \rebuttal{by} the sleeping position (left/right side). We demonstrated that the optimal rest position to minimize SBV can be \rebuttal{determined} by employing patient-specific simulation, which may have implications \rebuttal{for} the risk of LAA stasis and thus affect cardiovascular health in the medium and \rebuttal{long term}. 

Finally, some of the most common assumptions used in atrial simulations were tested to quantify their influence on atrial flow patterns\rebuttal{, and} non-Newtonian effects and the rigid-wall model were evaluated.


\section*{Acknowledgments}
This work was supported by \textit{Junta de Extremadura} and
FEDER funds under \rebuttal{Project} IB20105. We thank the \textit{Programa Propio - Universidad Polit\'{e}cnica de Madrid}, and the \textit{Ayuda Primeros Proyectos de Investigaci\'{o}n ETSII-UPM}. We also thank \textit{Programa de Excelencia para el Profesorado Universitario de la Comunidad de Madrid} for \rebuttal{their} financial support and the CeSViMa UPM project for its computational resources.

\section*{Conflict of interest}
The authors declare no conflict of interest.


\bibliographystyle{plainnat}
\bibliography{cas-refs}

\end{document}